\documentclass[twocolumn, numberedappendix]{aastex63}

\usepackage{amstext}
\usepackage{amsmath}
\usepackage{amssymb}
\usepackage{graphicx}

 \usepackage{enumerate}   

\newcommand{\aext}{{\bf a}_{\rm ext}}

\newcommand{\beq}{\begin{equation}}
\newcommand{\eeq}{\end{equation}}
\newcommand{\beqar}{\begin{align}}
\newcommand{\eeqar}{\end{align}}

\newcommand{\gloss}{\gamma_{\rm loss}}
\newcommand{\gw}{\gamma_{\rm wind}}
\newcommand{\ggrav}{\gamma_{\rm grav}}

\newcommand{\gL}{\gamma_{L_2}}

\newcommand{\lloss}{l_{\rm loss}}

\newcommand{\Ps}{\Phi_{\rm s}}
\newcommand{\ps}{p_{\rm s}}
\newcommand{\rhos}{\rho_{\rm s}}
\newcommand{\mdhi}{\dot M_{\rm hi}}
\newcommand{\mdlo}{\dot M_{\rm lo}}
\newcommand{\mde}{\dot M_{\rm est}}

\shorttitle{}

\shortauthors{}

\begin{document}

\title{Hydrodynamic Winds From Twin-Star Binaries}

\author[0000-0002-1417-8024]{Morgan MacLeod}
\affiliation{Center for Astrophysics $\vert$ Harvard \& Smithsonian, 60 Garden Street, Cambridge, MA, 02138, USA}
\email{morgan.macleod@cfa.harvard.edu}

\author[0000-0003-4330-287X]{Abraham Loeb}
\affiliation{Center for Astrophysics $\vert$ Harvard \& Smithsonian, 60 Garden Street, Cambridge, MA, 02138, USA}

\begin{abstract}
Stellar winds shape the evolution of stars through the loss of mass. In binary systems, they also shape the stars' evolution by modifying the orbit. In this paper, we use hydrodynamic simulations to study the emergence of nearly-isothermal winds from identical-twin binaries.  We vary the degree to which model stars fill their Roche lobes and the temperature of the wind. Initialized at rest on the stellar surfaces, winds accelerate away from the binary components through a sonic surface to supersonic outward velocities. In cases where the binary fills its Roche lobe, a shared subsonic region surrounds both components. We find that mass loss rates from close twin-star binaries are enhanced relative to the expectation from two single-object winds. This binary enhancement is best modeled as a function of the ratio of wind velocity to orbital velocity. Similarly, we find that the specific angular momentum with which winds emerge can vary between that of the binary components and that of the outer Lagrange points depending on the ratio of wind velocity to orbital velocity. Given that mass and angular momentum loss can be modeled as simple functions of wind velocity, our results may be broadly applicable to the evolution of close, equal-mass binaries. One particularly important potential application is to massive, close binaries which may be progenitors of binary black hole mergers through the chemically-homogeneous evolution channel. 
\end{abstract}


\section{Introduction}

Many stars exist in binary or multiple systems \citep{2013ARA&A..51..269D}. Among these, massive stars are particularly likely to be found in close pairs of similar-mass objects \citep{2012Sci...337..444S,2014ApJ...782....7D}. One such example is the over-contact pair VFTS352, which is composed of two approximately $29M_\odot$ components that fill their mutual Roche lobes \citep{2015ApJ...812..102A}. Sources like these are of significant interest as potential progenitor systems of binary compact object remnants that can inspiral and merge through the emission of gravitational radiation \citep{2016MNRAS.458.2634M,2016MNRAS.460.3545D,2016A&A...588A..50M,2016A&A...585A.120S}. In these systems, rotation assists internal mixing and leads much of the stellar hydrogen to be burned and incorporated into the compact core \citep{2009A&A...497..243D}, which later collapses. 

Stars ubiquitously lose mass via winds from their surfaces. Depending on the stellar type the wind acceleration mechanism and thermal properties can differ, and may depend on stellar metallicity or magnetic field \citep{1999isw..book.....L}. Wind mass loss significantly modifies the evolution of stars, and shapes the distribution of stellar remnant masses \citep[e.g.][]{2015MNRAS.451.4086S}. When stars are in binary or multiple systems, wind mass and angular momentum losses have the further effect of transforming the binary orbit.  The details of how much mass stars lose, and the specific angular momentum with which it is expelled from a binary system, are therefore crucial for understanding the details of how stellar winds affect evolving binary systems \citep[e.g.][]{1977MNRAS.179..265L,1983adsx.conf..343S,1991MNRAS.253....9T,1993ApJ...410..719B,2002MNRAS.329..897H,2018MNRAS.473..747C}. 

In close binaries, stellar winds interact with each other and with the combined gravitational effective potential of the pair. Although a variety of approaches have been used, some of the most informative models of this process come from hydrodynamic simulations. Much of this work has focused on the dynamics of stellar wind capture by companion objects \citep[e.g.][]{1990ApJ...356..591B,1991ApJ...371..684B,1991ApJ...371..696T,1993MNRAS.265..946T,1994ApJ...435..756B,1995ApJ...445..889B,1996MNRAS.280.1264T,2004A&A...419..335N,2005A&A...441..589J,2007ASPC..372..397M,2009ApJ...700.1148D,2012A&A...544A..59B,2013MNRAS.433..295H,2015A&A...575A...5C,2017MNRAS.468.3408D,2018MNRAS.475.3240E,2019A&A...622A.189E,2019A&A...629A.103S,2019MNRAS.488.5162X,2019MNRAS.490.3098T,2020MNRAS.493.2606B,2020A&A...637A..91E}, which is of particular observational importance for X-ray binaries in which the accretor is a compact object.  Other hydrodynamic modeling, also motivated by X-ray observations, has focused on the morphology and dynamics of colliding winds in binaries with pairs of wind-emitting stars, such as Wolf-Rayet binaries \citep[e.g.][]{1992ApJ...386..265S,1995ApJ...454L.145O,1998Ap&SS.260..243W,1998MNRAS.300..479P,2007ApJ...662..582L,2008MNRAS.388.1047P,2009MNRAS.396.1743P,2011MNRAS.418.2618L,2012A&A...546A..60L,2013A&A...560A..79L,2019A&A...625A..85V}. 

In the follow, we study the morphology of thermally-driven stellar winds, similar to the \citet{1958ApJ...128..664P} solar wind model, in close pairs of stars. We focus on equal-mass systems, with identical surface conditions, and we examine how the degree of Roche lobe occupancy (how close the system is to filling its Roche lobes) and the stellar-surface sound speed affect the emergent winds. We measure rates of mass and angular momentum loss from the binary system, and compare these to analytic predictions for single and binary systems. 

The paper is organized as follows. In Section \ref{sec:method}, we describe our hydrodynamic simulation method. In Section \ref{sec:results}, we describe our numerical results and discuss their interpretation. In Section \ref{sec:discussion}, we discuss potential limitations of our results and their broader applicability to astrophysical binaries with various wind-acceleration mechanisms. Finally, in Section \ref{sec:conclusion}, we conclude.

\section{Simulation Method and Models}\label{sec:method}

Our simulation model is composed of two identical stellar components, whose surfaces are in corotation with the binary orbit. We simulate the interaction of hydrodynamic winds launched from these surfaces. Our models are developed using the {\tt Athena++} code \citep{2020ApJS..249....4S}\footnote{Version 2019, https://princetonuniversity.github.io/athena}, which is an Eulerian (magneto) hydrodynamic code descended from {\tt Athena} \citep{2008ApJS..178..137S}.  

The total mass of the binary is $M$, the mass of the individual components is each $M_1 = M_2 = M/2$. The separation of their circular orbit is $a$. Our models are performed in units where $G = M = a = 1$. This implies that the unit velocity is $\sqrt{GM / a}=1$ and the unit time is $\sqrt{a^3 / GM}=1$, such that the orbital period is $P_{\rm orb} = 2\pi$. 

\subsection{Computational Method}

We solve the equations of inviscid gas dynamics in a frame of reference centered on the binary center of mass rotating with the binary orbital frequency $\Omega = \sqrt{GM/a^3}$. We employ a cartesian mesh, with nested levels of static mesh refinement surrounding the binary components. 

The conservation equations that we solve are 
\begin{subequations}\label{gaseq}
\begin{align}
\partial_t \rho  + \nabla \cdot \left( \rho {\bf v} \right) &= 0 , \\
\partial_t  \left( \rho {\bf v} \right) + \nabla \cdot \left( \rho {\bf v} {\bf v} + P {\bf I} \right)  &= - \rho \aext , \\
\partial_t E + \nabla \cdot \left[ \left( E+ P \right) {\bf v} \right] &= - \rho \aext \cdot \bf {v} ,
\end{align}
\end{subequations}
expressing  mass continuity, the evolution of gas momenta, and the evolution of gas energies. In the above equations, $\rho$ is the mass density, $\rho {\bf v}$ is the momentum density, and $E = \epsilon + \rho {\bf v} \cdot {\bf v} / 2$ is the total energy density with $\epsilon$ being the internal energy density. The pressure is $P$, $\bf I$ is the identity tensor, and $\aext$ is the acceleration associated with the binary and the rotating frame of reference.  These equations  are closed by an ideal gas equation of state, $P=\left(\gamma -1\right) \epsilon$, where $\gamma$ is the gas adiabatic index. 

The source terms of the binary's gravity and rotating reference frame are contained in the acceleration,
\beq
\aext = -\frac{GM_1}{|{\bf r_1} |^3} {\bf r_1}  -\frac{GM_2}{|{\bf r_2} |^3} {\bf r_2} - {\bf \Omega} \times {\bf \Omega} \times {\bf r} - 2 {\bf \Omega} \times {\bf v},
\eeq 
where $\bf r_i$ is the vectorial separation between a zone and the center of star i,  ${\bf \Omega} = (0,0,\Omega)$ is the vectorial orbital frequency, $r$ and $v$ are the position and velocity relative to the center of mass of the rotating reference frame (i.e. the position and velocity in the computational domain). Thus the ($x,y$) coordinates definine the binary orbital plane. There is no gravitational backreaction of the wind distribution on the binary. 

The boundary of the stellar surfaces is chosen by a value, $\Ps$, of the gravitational effective potential,
\beq
\Phi_{\rm eff} = -\frac{GM_1}{r_1} - \frac{GM_2}{r_2}  - {1\over 2} \Omega^2 R^2,
\eeq
where $R=\sqrt{x^2+y^2}$.  In regions close to $M_1$ or $M_2$ where $\Phi_{\rm eff} < \Ps$, we set the gas velocities to zero, and remove any acceleration terms. The density and pressure are set to fixed stellar ``surface" values of $\rho_{\rm s}$ and $p_{\rm s}$. We set $\rho_{\rm s} = 1 M/a^3$\footnote{Setting $\rho_{\rm s}$ has the effect of normalizing the mass loss rate. Because we do not include any backreaction on the binary orbit or any density-dependent heating or cooling physics, this parameter can be defined independently of the other choices of units.}, and the pressure based on the hydrodynamic escape parameter, 
\beq
\lambda = - \frac{\Ps}{c_{\rm s,s}^2},
\eeq
 where $c_{\rm s,s}^2 = \gamma p_{\rm s}/ \rho_{\rm s}$ is the squared sound speed of the stellar surface. Thus, 
 \beq
 \ps = - \frac{\rhos  \Ps }{ \gamma \lambda }.
 \eeq 
We specify the value of $\Ps$ by comparison to $\Phi_{\rm eff}$ at the $L_1$ Lagrange point, which, in the case of $M_1/M_2 = 1$ is located at the center of mass, 
\beq\label{PhiSurf}
\Ps = f_\Phi \Phi_{\rm eff} (L_1) = -2 f_\Phi ,
\eeq
where the second equality holds with our chosen masses and code units.

The simulation domain extends to $\pm 48a$ in each direction from the center of mass, located at the origin. The base mesh is composed of $256^3$ zones. We nest 5 additional levels of static mesh refinement interior to this base mesh, each containing $256^3$ zones, for example the region within $\pm 24$ is refined one level, within $\pm 12$ two levels, up to $\pm 1.5$ refined five times. The smallest zone sizes are cubes with sides of $3/256 \approx 0.0117$.  This mesh is decomposed into meshblocks of $16^3$ zones each. The outer boundary conditions in each direction are ``outflow", extending the conditions interior to the domain to the ghost zones.

\subsection{Models}

We run a suite of models varying $\lambda$ and $f_\Phi$. For the adiabatic index, we adopt the nearly-isothermal value $\gamma = 1.01$. We set $\lambda =2.5$, $5.0$ and $10.0$ with $f_\Phi =1$, $\sqrt{2}$, $2$, and $4$. In what follows we discuss these 12 parameter combinations of $\lambda$ and $f_\Phi$. 

\begin{table*}
\begin{center}
\begin{tabular}{ccccccccccccccc}
\hline
model & $\gamma$ & $\lambda$ & $f_\Phi$ & $p_{\rm s}$ & $c_{\rm s,s}$ & $E_{J,s}$ & ${\cal B}_{\rm s}$ & $\dot M$ & $\dot L$ & $\dot L_{\rm grav}$ & $\gloss$ & $\gw$ & $\ggrav$  & $v_{10}$ \\
\hline
A & 1.01 & 2.5 & 1.00 & 0.79 & 0.89 & -2.00 & 78.00 & -1.17e+00 & -3.85e-01 & 3.99e-02 & 1.31 & 1.45 & -0.14 & 2.89 \\
B & 1.01 & 2.5 & 1.41 & 1.12 & 1.06 & -2.83 & 110.31 & -5.15e-01 & -1.36e-01 & 1.24e-02 & 1.05 & 1.15 & -0.10 & 3.64 \\
C & 1.01 & 2.5 & 2.00 & 1.58 & 1.26 & -4.00 & 156.00 & -2.57e-01 & -6.56e-02 & 4.17e-03 & 1.02 & 1.09 & -0.07 & 4.56 \\
D & 1.01 & 2.5 & 4.00 & 3.17 & 1.79 & -8.00 & 312.00 & -7.61e-02 & -2.09e-02 & 2.90e-04 & 1.10 & 1.12 & -0.02 & 7.03 \\
E & 1.01 & 5.0 & 1.00 & 0.40 & 0.63 & -2.00 & 38.00 & -5.15e-01 & -2.15e-01 & 1.28e-02 & 1.67 & 1.77 & -0.10 & 1.76 \\
F & 1.01 & 5.0 & 1.41 & 0.56 & 0.75 & -2.83 & 53.74 & -2.10e-01 & -5.34e-02 & 1.32e-02 & 1.02 & 1.27 & -0.25 & 2.22 \\
G & 1.01 & 5.0 & 2.00 & 0.79 & 0.89 & -4.00 & 76.00 & -1.07e-01 & -2.10e-02 & 7.56e-03 & 0.79 & 1.07 & -0.28 & 2.82 \\
H & 1.01 & 5.0 & 4.00 & 1.58 & 1.26 & -8.00 & 152.00 & -2.78e-02 & -6.64e-03 & 3.39e-04 & 0.96 & 1.00 & -0.05 & 4.49 \\
I & 1.01 & 10.0 & 1.00 & 0.20 & 0.45 & -2.00 & 18.00 & -7.47e-02 & -6.41e-02 & -1.69e-02 & 3.43 & 2.53 & 0.91 & 1.15 \\
J & 1.01 & 10.0 & 1.41 & 0.28 & 0.53 & -2.83 & 25.46 & -9.72e-03 & -6.00e-03 & -2.75e-03 & 2.47 & 1.34 & 1.13 & 1.39 \\
K & 1.01 & 10.0 & 2.00 & 0.40 & 0.63 & -4.00 & 36.00 & -2.76e-03 & -1.14e-03 & 5.62e-05 & 1.66 & 1.74 & -0.08 & 1.71 \\
L & 1.01 & 10.0 & 4.00 & 0.79 & 0.89 & -8.00 & 72.00 & -7.87e-04 & -1.31e-04 & 2.51e-05 & 0.67 & 0.79 & -0.13 & 2.73 \\
\hline
\end{tabular}
\end{center}
\caption{Simulation parameters and results. Model parameters include: $\gamma$, the gas adiabatic index, $\lambda$, the hydrodynamic escape parameter, $f_\Phi$, the potential of the stellar surfaces relative to $\Phi_{L_1}$, $p_{\rm s}$ and $c_{\rm s,s}$ the pressure and sound speed of the stellar surface boundary condition, $E_{\rm J,s}$ and ${\cal B}_{\rm s}$, the surface Jacobi and Bernoulli parameters. Model results include $\dot M$, the total mass flux from the binary, $\dot L$, the total angular momentum flux away from the binary, $\dot L_{\rm grav}$, the portion of $\dot L$ attributable to gravitational torques on the binary by the wind distribution. $\gloss$ is the dimensionless specific angular momentum of the wind, while $\gw$ and $\ggrav$ are the portions of this angular momentum attributable to hydrodynamic and gravitational stresses, respectively. Finally, $v_{10}$ is the mean wind radial velocity at $r=10a$. Model results are best converted to astrophysical units and applied to astrophysical systems through the dimensionless specific angular momenta $\gloss$, $\gw$, and $\ggrav$, and through the comparison of $\dot M$ to analytic predictions, as approximated by equation \eqref{binenhance}.  }
\label{simtable}
\end{table*}

\section{Analysis Metrics}

\subsection{Mass and Angular Momentum Loss}
We measure the mass loss rate from the binary via a surface integral 
\beq\label{mdoteval}
\dot M = - \oint_S \rho \left( {\bf v} \cdot {\bf dA} \right),
\eeq
surrounding the binary. In practice, we perform this integral by summing across the outward faces, ${\bf dA}$,  of zones closest to a sphere, $S$,  surrounding the center of mass. We adopt a radius of $5a$ to define $S$ in what follows, but have confirmed that our results would be nearly constant for any choice of radius between $2a$ and $6a$. The flux of the $\hat z$-component of angular momentum through this surface is similarly measured as 
\beq\label{ldoteval}
 \dot L = {\bf \dot L} \cdot \hat z = - \oint_S \rho \left( {\bf r}\times {\bf v} \right) \cdot \hat z \left( {\bf v} \cdot {\bf dA} \right),
\eeq
where quantities are measured in the inertial (non-rotating) frame.  Because of the symmetry defined by the orbital plane, we find that the $\hat x$ and $\hat y$ components sum to zero to machine precision. 

Combining these two fluxes, we define the specific angular momentum with which gas is lost from the binary, 
\beq
\lloss = \frac{\dot L}{\dot M},
\eeq
and its dimensionless counterpart, scaled to the specific angular momentum of the binary,
\beq
\gloss = \frac{l_{\rm loss} }{ l_{\rm bin} },
\eeq
where $l_{\rm bin} = L/M = 1/4$. Thus, $\dot L = \dot M l_{\rm loss}$ and $L = M l_{\rm bin}$. 

Angular momentum is acquired by the wind through a combination of the rotational motion of the binary about the center of mass, hydrodynamic stresses, and gravitational stresses. To understand this decomposition, it can be useful to separate these components. 

The gravitational stress on the fluid within the volume enclosed by $S$ implies a rate of change of the binary's angular moment that is equal and opposite the torque on the wind from the binary,
\beq
\dot L_{\rm grav} =  \dot {\bf L}_{\rm grav} \cdot \hat z =   \int  \sum_{i=1,2}\left( {\bf r_i} \times {\bf f}_{\rm grav,i}  \right) \cdot \hat z \ \rho dV,
\eeq
where ${\bf r_i}$ is the position of binary component $i$, and
\beq
{\bf f}_{\rm grav,i} =  \frac{GM_i}{ |{\bf r_i}|^3} {\bf r_i}  ,
\eeq
is the gravitational force of the binary on the wind per unit mass. Without loss of generality, we choose the orientation of the along the x-axis, so that the integrand can be simplified to
\beq
\sum_{i=1,2} \left( {\bf r_i} \times {\bf f}_{\rm grav,i}  \right) \cdot \hat z  =  x_1 \frac{GM_1}{ |{\bf r_1}|^3} y  + x_2 \frac{GM_2}{ |{\bf r_2}|^3} y,
\eeq
where $x_1 = -1/2$ and $x_2 = 1/2$. 

We define 
\beq
l_{\rm grav} = -\frac{\dot L_{\rm grav}}{\dot M}
\eeq
to describe the specific angular momentum imparted to the wind by torques from the binary, and also define
\beq\label{ggrav}
\ggrav = \frac{l_{\rm grav} }{l_{\rm bin} }.
\eeq
Finally, we distinguish the portion of wind specific angular  momentum not arising from gravitational torques as
\beq
l_{\rm wind} = l_{\rm loss} - l_{\rm grav}
\eeq
and 
\beq\label{gw}
\gw = \frac{l_{\rm wind}}{l_{\rm bin}}
\eeq
such that the total $\gloss$ is the sum of the hydrodynamic (wind) and gravitational components, $\gloss = \gw + \ggrav$.

\subsection{Wind Properties}

In addition to typical hydrodynamic properties, we define two useful characteristics of the wind, the Jacobi and Bernoulli parameters. Each of these is invariant in certain circumstances --  
and analyzing these properties allows us to decompose the hydrodynamic and gravitational stresses acting on the wind. 

The Jacobi parameter is,
\beq\label{EJ}
E_J = {1 \over 2} v_{\rm rot}^2  + \Phi_{\rm eff},
\eeq
where $v_{\rm rot}$ is the magnitude of the velocity in the frame rotating with the binary. In the restricted three-body problem \citep{1999ssd..book.....M}, test particles in the binary potential follow trajectories of constant Jacobi parameter. Constant Jacobi parameter along wind trajectories thus indicates that material is expanding freely along ballistic trajectories. Variations in Jacobi parameter indicate the importance of hydrodynamic stresses. 

The Bernoulli parameter of material in the wind is 
\beq\label{Bern}
{\cal B} = {1\over 2} v^2 + \Phi + h,
\eeq
where $h$ is the fluid enthalpy,
\beq
h = \frac{\gamma}{\gamma-1} \frac{P}{\rho}.
\eeq
For material at rest in the rotating frame (as is the case for the surface boundary conditions of our model stars) ${\cal B_{\rm s}} = E_{J,s} + h = \Phi_{\rm s} + h$. These values are tabulated in Table \ref{simtable}. The Bernoulli parameter is constant along fluid streamlines, such as a freely-expanding wind that does not self-intersect \citep[e.g.][]{2006iafd.book.....T}.

\subsection{Binary Orbital Evolution}
We begin with the expression for the orbital angular momentum of a binary system, 
 $L= M_1 M_2 /  M \sqrt{GMa}$. 
In our case $M_1 = M_2  = M/2$ , so 
\beq
L = {1\over 4} \sqrt{G M^3 a}. 
\eeq
By differentiating $L$ with respect to time, then dividing by $L$, we find,
\beq
\frac{\dot a}{a} = \frac{2\dot L}{L} -  \frac{3\dot M}{M}
\eeq
measuring $\dot M$ and $\dot L$ allow us to estimate the separation evolution of the circular orbit.  

Substituting in the definition of $\gloss$, the orbit evolution equation can be rewritten 
\beq\label{adot}
\frac{\dot a}{a} = \left( 2 \gloss -3 \right)  \frac{\dot M}{M},
\eeq
where $\dot M < 0$, so $\gloss > 3/2$ implies that $\dot a <0$, while  $\gloss < 3/2$ implies that $\dot a > 0$. This can be re-written, 
\beq
\frac{d\ln a}{d\ln M} = 2 \gloss -3 ,
\eeq
in terms of the orbital separation change per unit mass change. Integrated over some change in binary mass, the ratio of final to initial separation $a_{\rm f}/a_{\rm i}$ depends on the ratio of final binary mass to initial mass, $M_{\rm f}/M_{\rm i}$,
\beq
\frac{a_{\rm f}}{a_{\rm i}} = \left( \frac{M_{\rm f}}{M_{\rm i}} \right)^{2 \gloss -3}.
\eeq
Thus, when $\gloss =1$, this takes on the simple form, $a_{\rm f}/a_{\rm i} = M_{\rm i}/M_{\rm f}$.

\section{Results}\label{sec:results}

\subsection{Twin Wind Morphology}

\begin{figure*}[htbp]
\begin{center}
\includegraphics[width=0.45\textwidth]{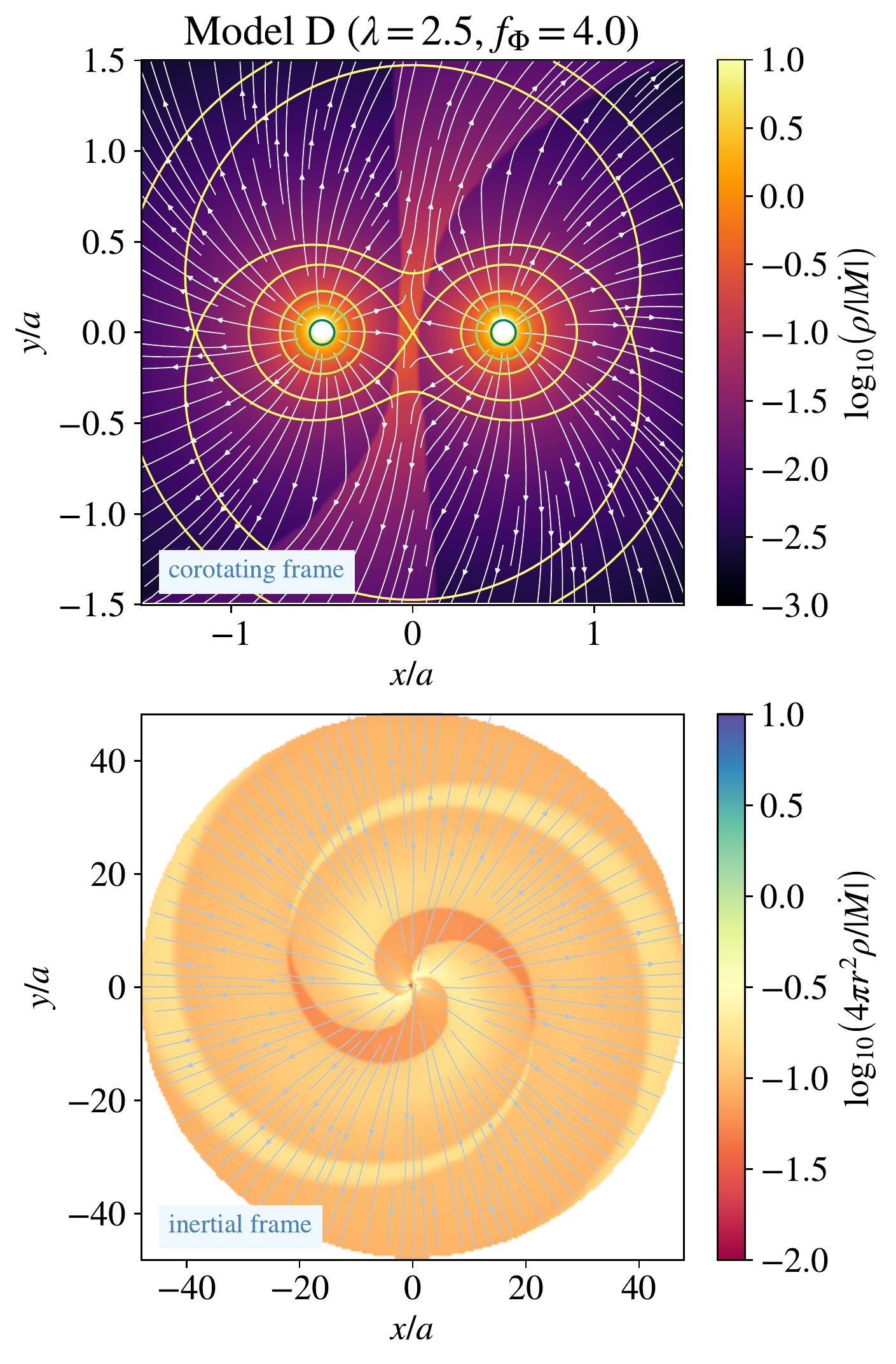}
\includegraphics[width=0.45\textwidth]{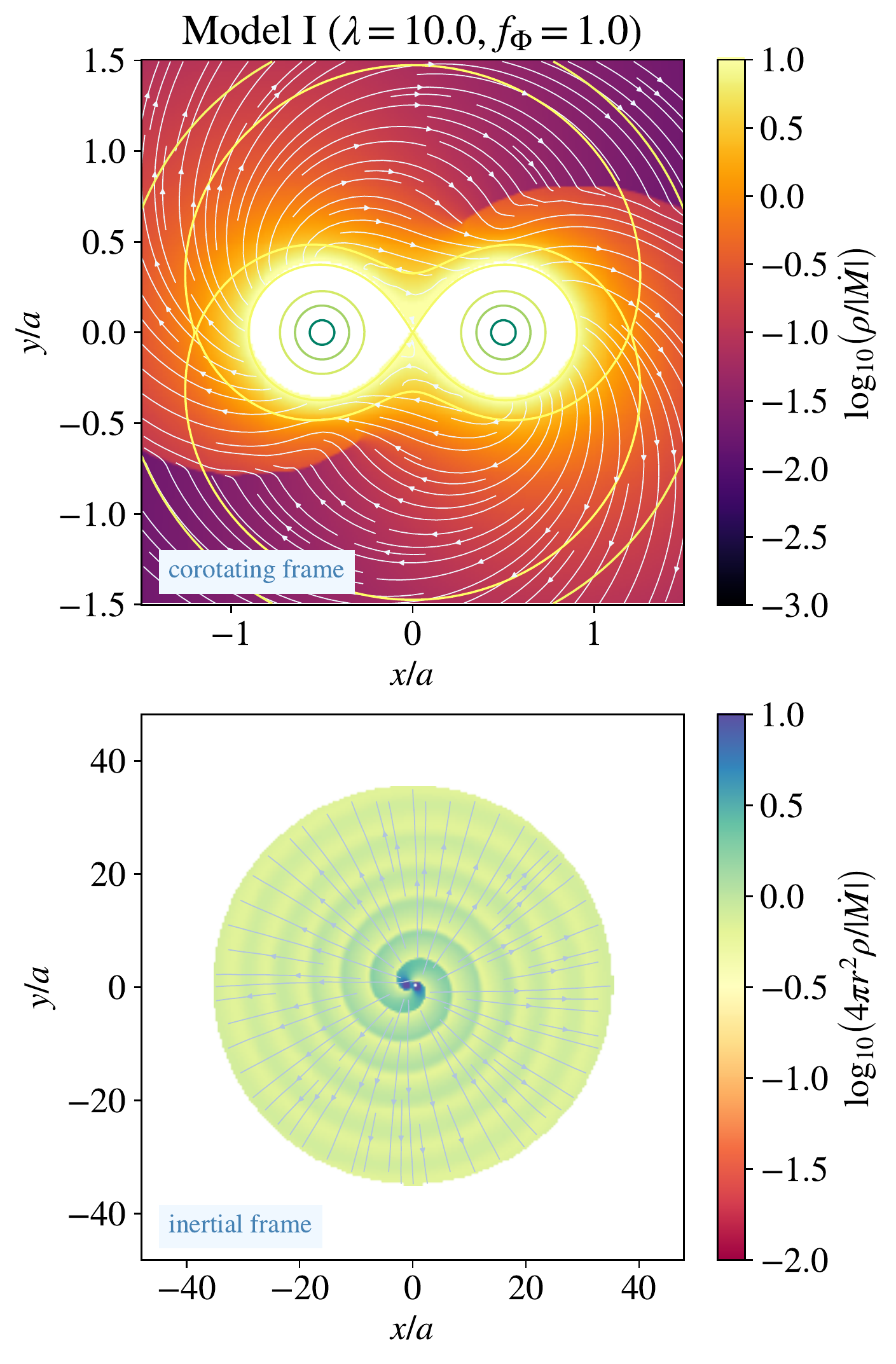}
\caption{Slices of wind density normalized by mass loss rate (upper panels) and with the approximate $r^{-2}$ scaling removed (lower panels). Streamlines in the upper panels are plotted in the corotating frame, while streamlines in the lower panels are shown in the inertial frame. Winds from the two components join and interact near the binary, before forming a largely-spherical outflow with spiral disturbances on larger scales.  The complete figure set (12 images) is available at \url{https://github.com/morganemacleod/TwinWinds}.  }
\label{fig:rho}
\end{center}
\end{figure*}

The thermal winds in our model are accelerated from rest on the surfaces of the binary components by the pressure gradient established as the wind expands into the surrounding space.  Figure \ref{fig:rho} shows slices of density in the binary equatorial plane (the figure set online shows each of the Models in Table \ref{simtable}). The upper panels show density, normalized by the mass loss rate, in the region in the vicinity of the binary components with flow streamlines in the corotating frame. In the lower panels, we plot $4\pi r^2 \rho / |\dot M|$, to visualize deviation from a spherically-expanding constant radial velocity wind.  In this panel, velocity streamlines are shown in the non-rotating inertial reference frame. 

The wind density structures in the vicinity of the binary depend greatly upon the degree of contact of the binary components, parameterized by $f_\Phi$, and the hydrodynamic escape parameter $\lambda$.  The influence of these parameters may be compared in Figure \ref{fig:rho}. For Model D, in which $\lambda = 2.5$ and $f_\Phi = 4$, the relatively high surface sound speed leads to rapidly expanding, supersonic winds. The separation of the binary relative to the component sizes leads the winds to establish separately prior to colliding in an interaction region. The rotation of the binary system imparts a spiral shape to this collision sheet \citep{2007ApJ...662..582L}. By contrast, Model I, for which $\lambda = 10$ and $f_\Phi=1$, exhibits a dense circulating layer surrounding the binary. Material trails away from the binary system along the leading edges of the rotating pair, passing through the outer, $L_2$ and $L_3$ Lagrange points (which have identical potential for our equal-mass case). Other models exhibit behavior intermediate between these extremes, with winds superimposing to form spiral structures emanating from the binaries' vicinity. 

In examining the large-scale density and kinematic structures, we find that for each of our models  the wind expands nearly radially in all directions (with $v_\phi \ll v_r$), and that wind densities reflect this nearly-spherical expansion with an approximately $\rho \propto r^{-2}$ density structure.  Deviations from this baseline behavior of a spherical wind represent differences relative to spherical expansion at constant velocity. Figure \ref{fig:rho} shows that the binary's motion imparts spiral waves on the expanding winds. The winding angle of these waves depends on the normalization of the expansion velocity relative to orbital velocity (high velocity winds will be loosely wound, expanding further in each orbital cycle). The overall normalization of $4\pi r^2 \rho / |\dot M|$ for each model reflects the inverse of the wind expansion velocity -- higher velocity winds have lower densities at the same mass loss rate.  Spiral waves impart an approximately order of magnitude variation in local density in the orbital plane over the spherical-mean. These fluctuations are quickly reduced out of the orbital plane -- near the poles the density profile is a smooth $r^{-2}$ pattern. We note that wind interaction regions lead to changes of wind angular velocity as angular momentum is redistributed and averaged among the interacting winds. However, because the velocity is primarily radial, these appear as only slight deviations in the streamlines in even the most extreme cases, as seen in the lower panel for Model D.

\subsection{Wind Acceleration and Velocity}

\begin{figure}[htbp]
\begin{center}
\includegraphics[width=0.45\textwidth]{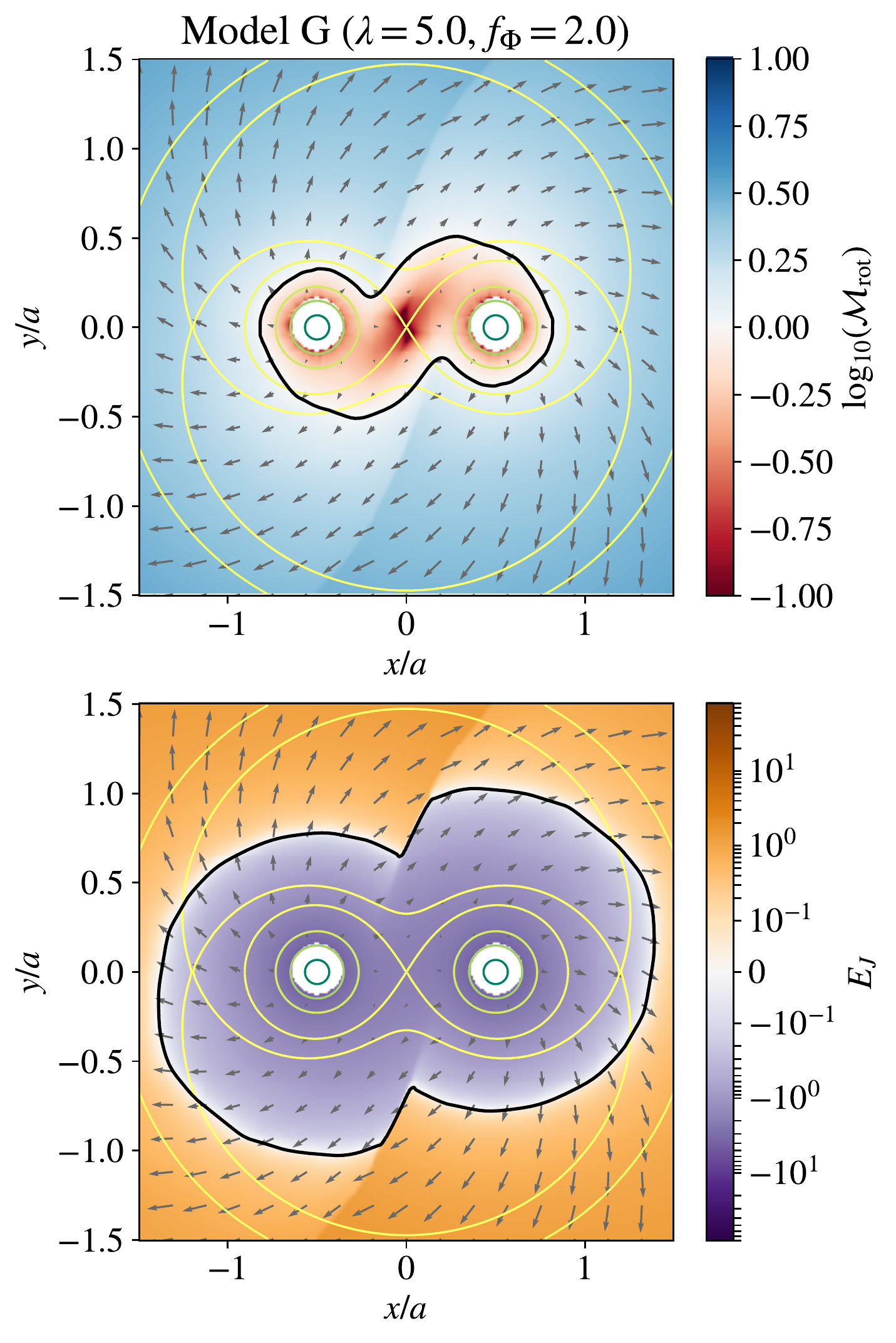}
\caption{Slices of wind Mach number (upper panels) and Jacobi parameter (lower panels) at $z=0$ in the corotating frame. The black contours shows the ${\cal M}_{\rm rot}=1$ sonic surface, and the $E_J = 0$ surface. The sonic transition happens closest to the surfaces of the binary components for $\lambda = 2.5$, and furthest for $\lambda=10$. The initial Jacobi parameter on the surface of the binary is $E_{J,0}=-2 f_\Phi$. Because $E_J$ is constant for collisionless motion in the binary potential, increases in $E_J$ represent the acceleration of the wind by pressure gradients.  The complete figure set (12 images) is available at \url{https://github.com/morganemacleod/TwinWinds}. }
\label{fig:machEJ}
\end{center}
\end{figure}

This discussion of the interaction and superposition of the binary components' winds reveals the critical role of the winds' emergent velocity.
Thermal winds accelerate due to radial pressure gradients from near-rest through a critical point, at which the expansion velocity equals the sound speed, eventually reaching supersonic radial expansion.
In Figure \ref{fig:machEJ}, we analyze slices through the binary systems' orbital planes (Figure set online shows each model). The upper panels show Mach number in the corotating frame, the lower panels show Jacobi parameter. Velocity vectors are overplotted relative to the corotating frame. Mach number and Jacobi parameter provide a window into the role that pressure gradients play in accelerating the wind away from the binary.

In the Mach number panels of Figure \ref{fig:machEJ}, we see the transition from subsonic outflow near the binary components to supersonic outflow with increasing radius -- we mark the critical transition at ${\cal M}_{\rm rot} =1$ with a contour.  For spherical, isothermal winds, the critical point radius is $r_{\rm sonic} = GM / 2 c_{\rm s}^2 = (\lambda /2 ) R_{\rm s}$, where $R_{\rm s}$ is the radius of the stellar surface.  This relation implies that we expect a larger subsonic outflow region for larger $\lambda$, because the shallower pressure gradients in these cases accelerate the wind more slowly. For some of the model parameter space, there are individual sonic transitions surrounding each binary component. This occurs for low $\lambda$, when the sonic radius becomes closer to the object radius, and for larger $f_\Phi$, when the objects fill smaller fractions of their Roche lobes. However, other cases, such as Model G, in which $\lambda = 5$ and $f_\Phi=2$, the sonic surface surrounds and encloses both binary components. For $\lambda = 2.5$, we observe that the sonic surface generally lies just outside of our surface boundary condition,  is joined across both binary components only for the contact case of $f_\Phi = 1$. For $\lambda = 5$, when $f_\Phi \leq 2$ the sonic surface is joined, while for $\lambda = 10$, the sonic surface is joined across the binary components for all $f_\Phi$ studied.  These findings are consistent with the simple estimate of joined sonic surfaces when $R_{\rm s}/ R_{\rm Roche} \gtrsim 2 / \lambda$, or in terms of our model parameters, $f_\Phi \lesssim \lambda / 2$. 

The variation of the Jacobi parameter in model slices directly illustrates the role of pressure gradients in accelerating the wind. As we tabulate in Table \ref{simtable}, the surface Jacobi parameter, equation \eqref{EJ}, of the wind is equal to the effective potential of the surface.  Ballistic, collisionless motion in the binary's gravity occurs at constant Jacobi parameter. By contrast, the increasing Jacobi parameter seen in the slices of Figure \ref{fig:machEJ} is the result of the collisional nature of the fluid, and the degree to which pressure gradients add kinetic energy to the flow.
We mark the surface of zero Jacobi parameter in Figure \ref{fig:machEJ}, this represents the transition at which a collisionless particle is unbound relative to the binary system, ignoring the gas' internal energy. Under different model parameter variations, this $E_J = 0$ surface occurs at varying distances from the binary components. As for the sonic surface, at low $\lambda$ and high $f_\Phi$ this transition separately surrounds the two binary components, while for higher $\lambda$ and lower $f_\Phi$, this energetic transition surface surrounds the binary (for some of the parameter variations it occurs outside the $\pm 1.5a$ box of Figure \ref{fig:machEJ}). 

The gradients of $E_J$ are strongest for the highest sound speed winds, however we note that outflowing streamlines trace increasing $E_J$ in all cases. Streamlines along which Jacobi parameter is relatively constant, such as the circulating flow in the subsonic region close to the binary (in Model I, for example) indicate regions where circulation at constant energy is occurring mostly under the influence of the binary gravitational potential. 
By contrast with the increases observed in $E_J$, we note that the wind Bernoulli parameter, equation \eqref{Bern}, is observed to be constant within the winds to with approximately 2\% of the surface Bernoulli parameter tabulated in Table \ref{simtable}. Together with our discussion of $E_J$, this indicates that the acceleration of the wind away from the binary is due to the energy associated with the gas enthalpy, $h$, which results in the increasing velocity of the wind along outflowing streamlines. 

\begin{figure}[tbp]
\begin{center}
\includegraphics[width=0.49\textwidth]{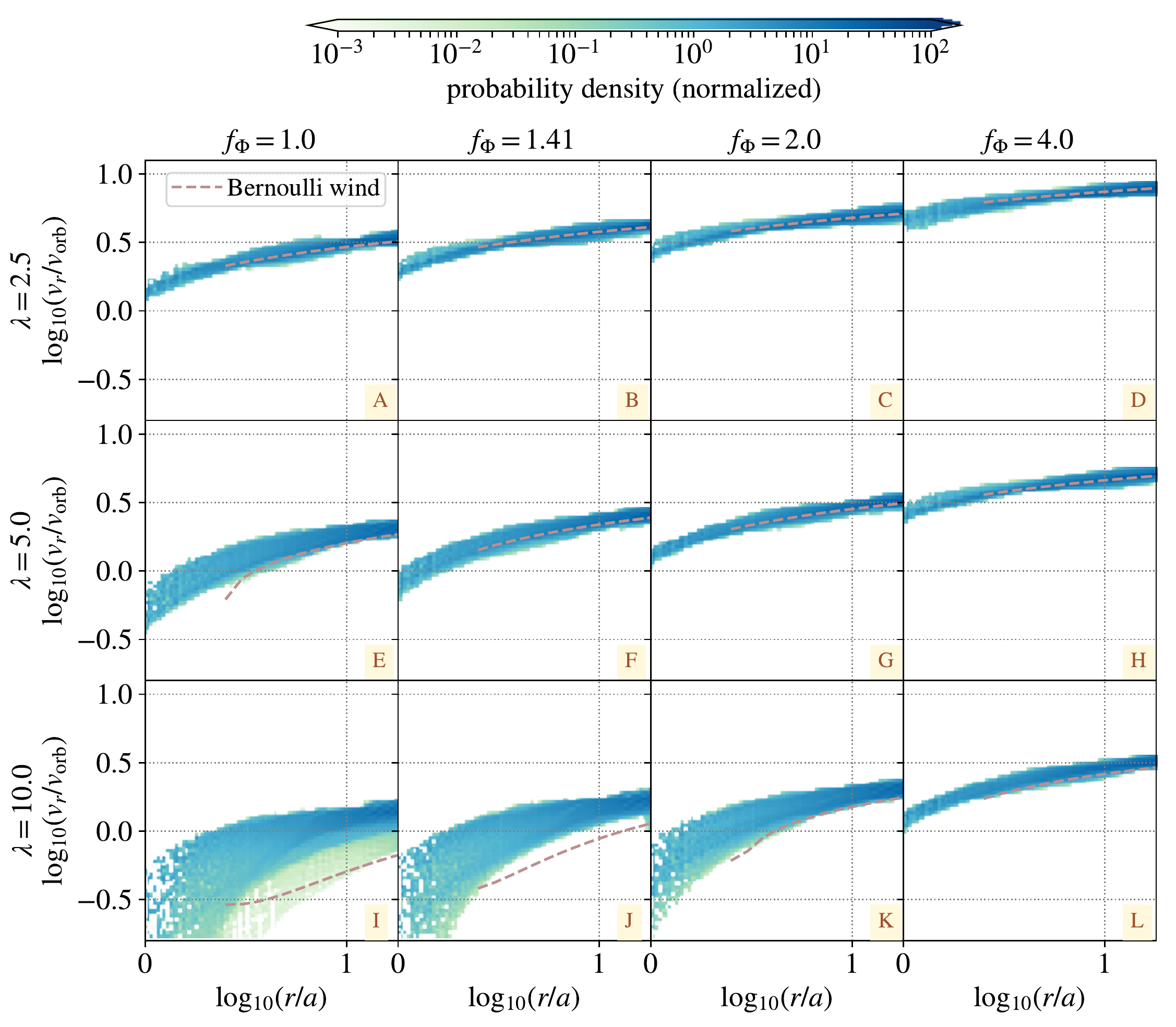}
\caption{ Radial velocity of wind with $r<18a$ of the binary center of mass.  The phase plot is colored by the normalized probability distribution function of wind mass in the $r$ versus $v_r$ plane. In each case we compare to the velocity profile of polytropic winds from a single object with mass equal to the binary mass, and $\dot M$ equal to that of the binary simulation, labeled ``Bernoulli wind".   }
\label{fig:vr}
\end{center}
\end{figure}

Figure \ref{fig:vr} examines the development of radial velocity in the wind more quantitatively. Within a spherical volume of $r<18a$, we sample each zone and plot its radial velocity, $v_r$, relative to $r$. The resulting velocity--radius phase plot is colored by relative mass per pixel. In each case we see that the wind is accelerated steeply over $r \lesssim 3 a$, and then more gently as the wind continues to expand. A purely isothermal spherical wind will continue to grow in radial velocity over whatever scale the isothermal temperature is maintained. Our model winds with $\gamma=1.01$ have finite, but large, asymptotic velocity. However, it is worth noting that the radial velocity is close to constant outside of the primary acceleration region that corresponds to the subsonic regions close to the binary in Figure \ref{fig:machEJ}. In Figure \ref{fig:vr}, we compare to the velocity profiles of corresponding spherical polytropic winds, following the formalism outlined in Appendix \ref{sec:wind}. We solve the Bernoulli equation to model the wind with mass flux $\dot M$ equal to that of the binary simulation from an object with central mass $M$. These solutions are marked ``Bernoulli wind" in Figure \ref{fig:vr}. We observe that the binary imparts some variation in the wind radial velocity at a given radius, but that the overall trend tracks that of the spherical wind, especially for $\lambda =5$ and $\lambda=2.5$. For $\lambda=10$, we observe that the wind radial velocity in the binary simulations is larger than that predicted by the Bernoulli model. 

\begin{figure}[tbp]
\begin{center}
\includegraphics[width=0.45\textwidth]{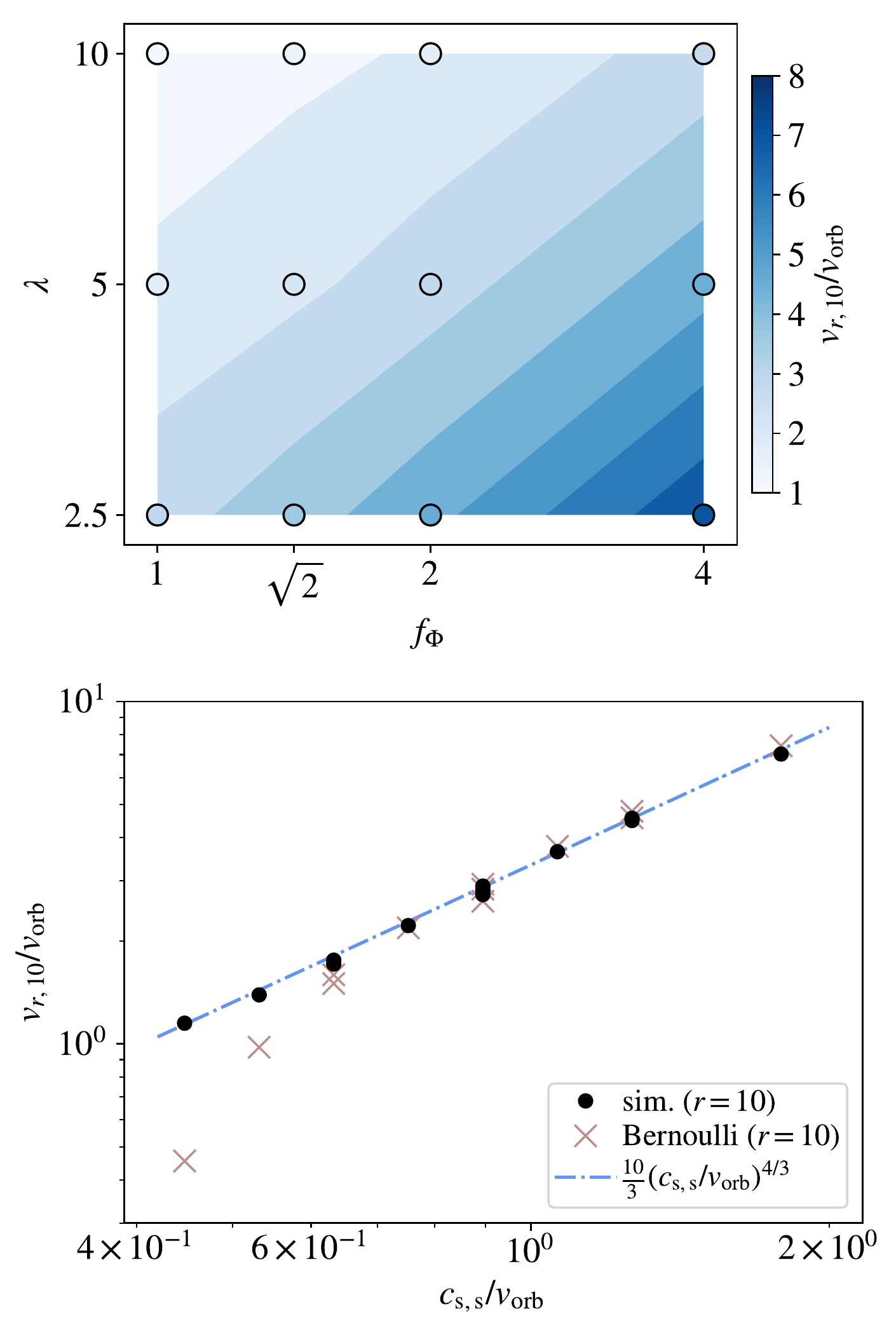}
\caption{  Wind radial velocity at $r=10a$, as a function of surface sound speed, $c_{\rm s,s}$, both normalized by the binary's orbital speed, $v_{\rm orb}$. We compare to our solution of the Bernoulli equation for a spherical wind with mass flux $\dot M$ from an object of mass $M$, labeled ``Bernoulli wind". We observe that the spherical wind largely predicts the velocities at $10a$, except at lower $c_{\rm s,s}$, where there is some asymptotic behavior in the simulations with $v_{r,10}\sim v_{\rm orb}$,  due to the orbital motion imparting radial velocity to the wind. }
\label{fig:v10}
\end{center}
\end{figure}

Across binary parameters, we observe significant trends in the velocity to which the winds are accelerated. We measure the wind velocity at $r=10a$ by taking the mass-weighted mean wind speed in a spherical shell $9.5 < r/a < 10.5$, and denote the result $v_{r,10}$.   Beyond this, the velocity of our polytropic winds continues to expand slowly \citep[and indeed an isothermal wind's velocity increases across as large a region as the isothermal temperature is maintained, cf.][]{1958ApJ...128..664P}.   Figure \ref{fig:v10} shows the dependence of $v_{r,10}$ on the model parameters $\lambda$ and $f_\Phi$. We see that the highest velocity winds arise for high $f_\Phi$ and low $\lambda$. Because $c_{\rm s,s}^2 = - \Phi_{\rm s} / \lambda = - \Phi_{L_1} f_\Phi / \lambda$, clearly there is a trend that relates the wind velocity to the surface sound speed.

The lower panel of Figure \ref{fig:v10} compares the surface sound speed to the resultant wind velocity.  These initial sound speeds are listed in Table \ref{simtable}. We note that the wind speed is well described by a power-law dependence,
\beq
\frac{v_{\rm r,10} }{v_{\rm orb} } \approx \frac{10}{3}  \left( \frac{c_{\rm s,s}}{v_{\rm orb}} \right)^{4/3},
\eeq
which we additionally plot in Figure \ref{fig:v10}. 
Finally, we compare the trends in $v_{r,10}$ as a function of $c_{\rm s,s}$ by solving the Bernoulli equation for a spherical wind for each of the model parameters. The Bernoulli wind model accurately predicts the wind velocities for most of the simulations, except those with the lowest surface sound speeds. These models with low $c_{\rm s,s}$, exhibit higher radial velocities in the simulation which appear to asymptote to  $v_{r,10}\sim v_{\rm orb}$, implying that the binary's orbital motion plays a significant role in imparting radial velocity to these winds. 

\subsection{Mass Loss Rate}

\begin{figure}[tbp]
\begin{center}
\includegraphics[width=0.45\textwidth]{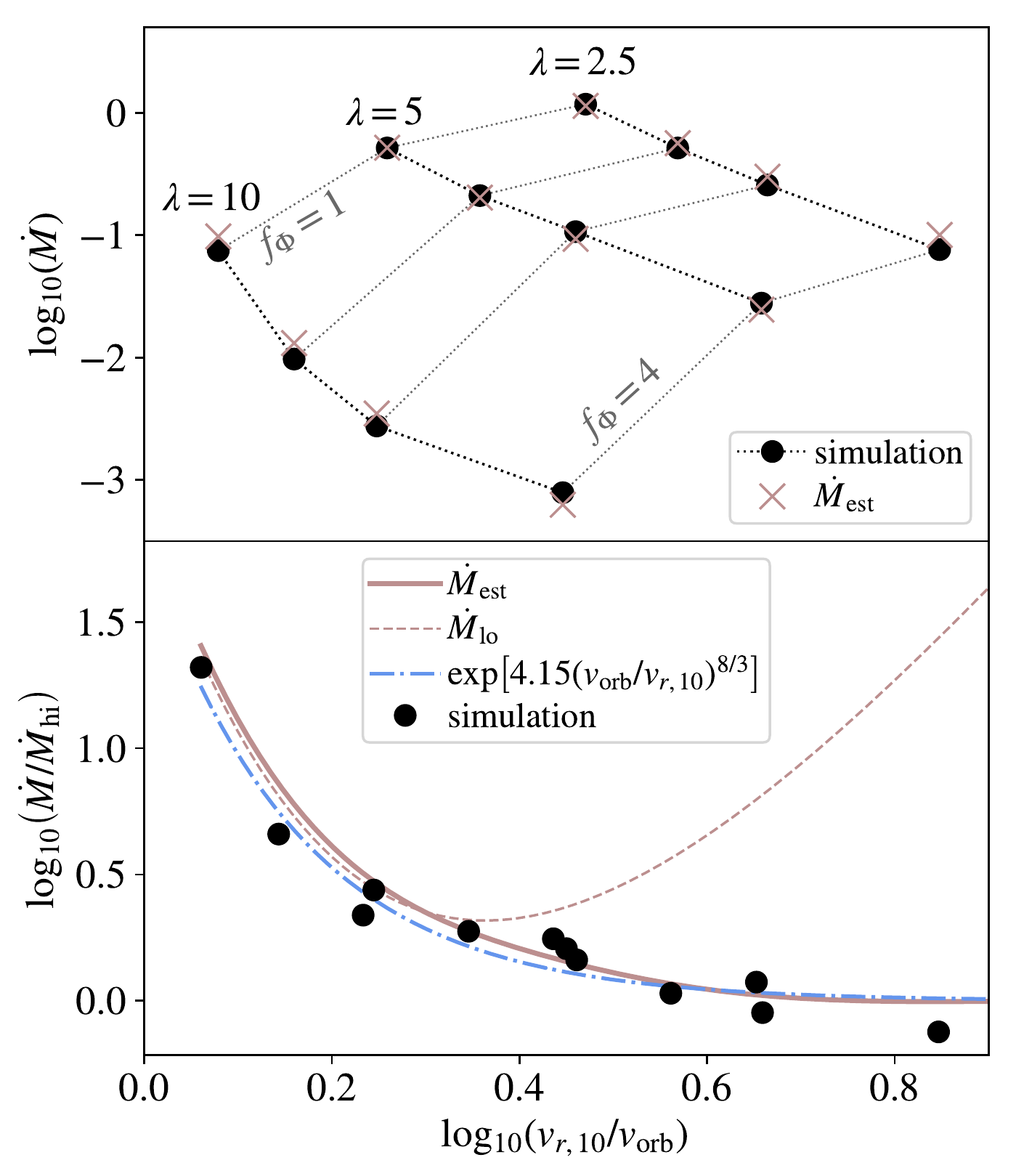}
\caption{Mass loss rates from simulated twin-star binaries as a function of wind velocity, $v_{r,10}$ in units of $v_{\rm orb}$.  In the upper panel, we show simulation $\dot M$ in code units compared to $\mde$, equation \eqref{mde}. The highest mass loss rates occur for the combination of low $\lambda$ and $f_\Phi$. In the lower panel, we normalize $\dot M$ by $\mdhi$, twice the expected single-object mass loss rate. At low wind velocities relative to the orbital velocity (equivalently for close binaries at fixed wind velocity), the mass loss rate from the binary is enhanced relative to the non-binary equivalent.   }
\label{fig:mdot}
\end{center}
\end{figure}

We measure the mass loss rate from the binary via the surface integral of equation \eqref{mdoteval}. The resulting mass loss rates are tabulated in Table  \ref{simtable}. In all cases, mass is flowing away from the binary in a steady-state wind.

In Figure \ref{fig:mdot}, we show the resulting mass loss rates as a function of wind velocity, labeling groups of $f_\Phi$ and $\lambda$ model parameters (upper panel). At constant $f_\Phi$, we observe that when the wind velocity (or surface sound speed) is larger (smaller $\lambda$), $ |\dot M|$ increases.  By contrast, when $f_\Phi$ increases at constant $\lambda$, the increased wind velocity implies decreasing $ |\dot M|$, implying that the deeper potential well implied by larger $f_\Phi$ leads to more difficult escape for winds of a given velocity. 

To some extent, these trends align with our analytic understanding of hydrodynamic winds. To establish a baseline for comparison, we define several estimates of the mass loss rate based on the binary parameters and surface sound speed, which are derived in Appendix \ref{sec:mdotest}. We define:
 \begin{enumerate}[i)]
 \item  $\mdhi$, twice the estimated isothermal mass loss rate from two individual objects of mass $M/2$, which have surface sound speed $c_{\rm s,s}$ and surface density $\rhos$. This estimate of the mass loss rate is derived by computing $2 \times 4\pi r_{\rm sonic}^2 \rho_{\rm sonic} c_{\rm s,s}$. As we will discuss, $\mdhi$ applies in the regime of high-velocity winds that escape with little interaction with the binary potential.
 \item  $\mdlo$, the outflow from the vicinity of the outer saddle points of the gravitational potential, $L_2$ and $L_3$, by estimating $2 \times \rho_{L_2} A_{\rm L_2} v_{\rm L_2}$, where $v_{\rm L_2} \sim c_{\rm s,s}$, and the estimation of the other terms is discussed in detail in the Appendix. This estimate is applicable in the case of lower-velocity winds that interact strongly with the binary potential. 
 \end{enumerate}

 We find that the following formula interpolates smoothly between the expected behaviors at high and low sound speed,
\begin{align}\label{mde}
\mde &\approx - \pi \left[  \left( \frac{ G^2 M^2 }{2 c_{\rm s,s}^3} \right)^{-1}   + \left( \frac{2  c_{\rm s,s}^3  a^3 }{GM}\right)^{-1}  \right]^{-1} \nonumber  \\ 
                     &\times \rhos \exp\left[ - \lambda + \sqrt{  \left( -\frac{\Phi_{L_2}}{c_{\rm s,s}^2 } - \frac{1}{2}  \right)^2   + \left(\frac{3}{2}\right)^2  } \right].
\end{align}
This estimating formula is plotted over the simulation data in Figure \ref{fig:mdot}. This comparison shows that $\mde$  qualitatively captures the trends in $\dot M$ with binary parameters as well as the overall normalization of the mass loss rate. 

In the lower panel of Figure \ref{fig:mdot}, we compare the mass loss rates derived from the binary simulations to that estimated for two single objects, $\mdhi$, as a function of wind velocity. This ratio expresses the enhancement in mass loss rate that results from the binary potential at low wind velocities. Furthermore, we observe that the results collapse to a single velocity dependent relationship under this normalization. Through these comparisons, we observe that $\mdlo$ captures the low velocity behavior, but diverges at high velocity, while $\mde$ provides a reasonable approximation across the wind velocity range. The ratio of $\mde/ \mdhi$ is
\beq \label{mderatio}
\frac{\mde}{\mdhi} \approx \frac{ \exp\left[ - \frac{3}{2} + \sqrt{  \left( -\frac{\Phi_{L_2}}{c_{\rm s,s}^2 } - \frac{1}{2}  \right)^2   + \left(\frac{3}{2}\right)^2  } \right]}{1  + \frac{1}{4} \left( \frac{v_{\rm orb} }{c_{\rm s,s}}\right)^{6} },
\eeq
which provides an analytic expression for the enhancement in the hydrodynamic wind from the binary solely as a function of its properties and surface sound speed. 

Finally, the fact that the binary enhancement is a function of wind velocity indicates that we may be able to express the binary enhancement in wind mass loss rate as a function of the general property of wind velocity. 
Inspired by the functional form of equation \eqref{mderatio}, we fit $\dot M / \mdhi$ to an exponential function, and find that,
\begin{align}\label{binenhance}
\frac{\dot M}{\mdhi} &\approx \exp \left[ 2 \left( \frac{-\Phi_{L_2}}{v_{r,10}^2}\right)^{4/3}  \right], \nonumber \\
&\approx \exp \left[ 4.15 \left( \frac{v_{\rm orb}}{v_{r,10}}\right)^{8/3}  \right], 
\end{align}
provides a good description of the enhancement in $\dot M$ due to the fact that the objects are in a close binary. This function is shown with a dot-dash line in Figure \ref{fig:mdot}.  For example, for a pair of stars with fixed wind velocity, equation \eqref{binenhance} describes how the total mass loss rate changes as the binary separation (and thus $v_{\rm orb}$) changes. 

\begin{figure}[tbp]
\begin{center}
\includegraphics[width=0.4\textwidth]{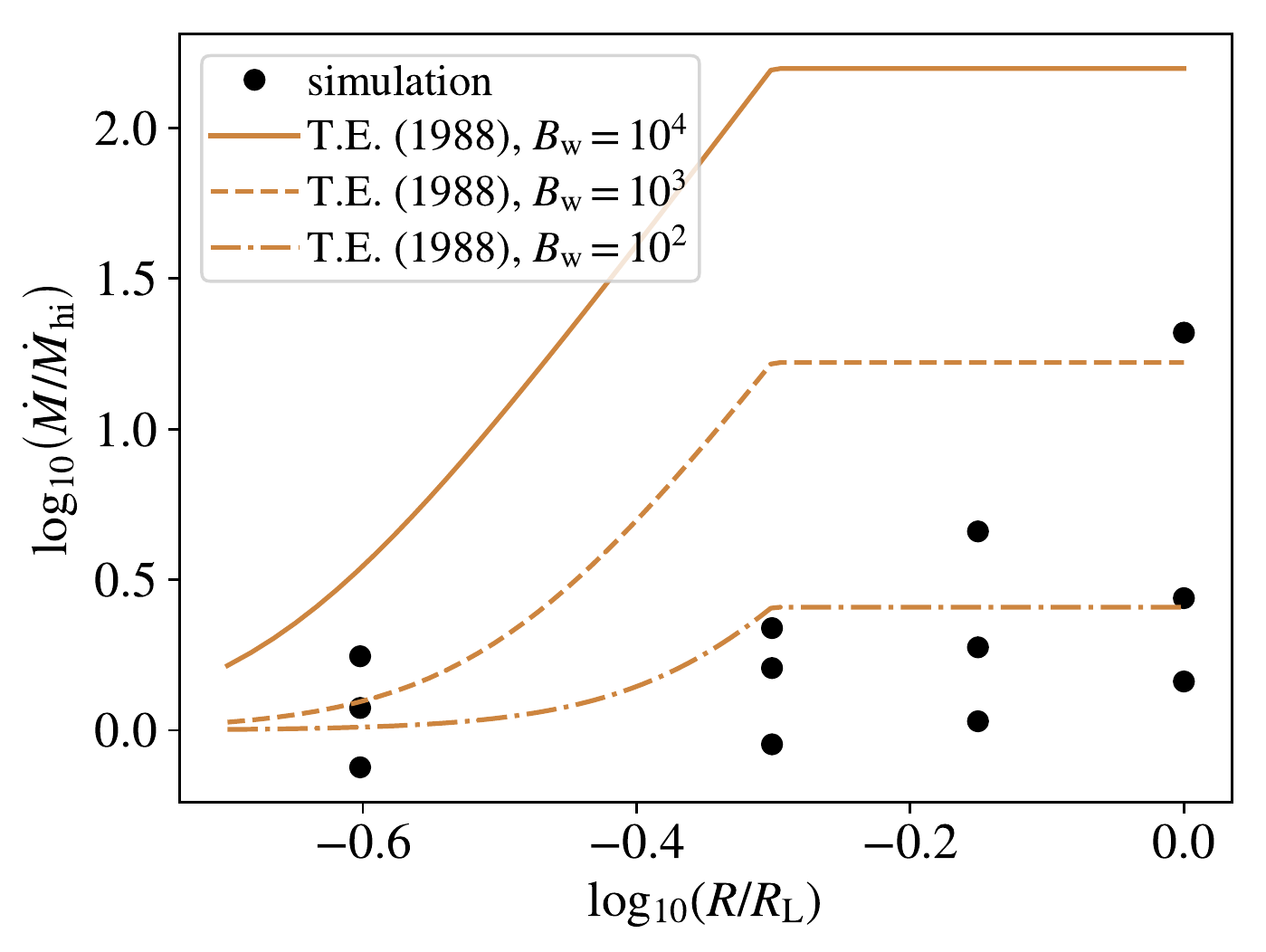}
\caption{ Comparison of simulated enhancement in mass loss rates compared to the single-object prediction, as a function of the ratio of the stellar radius to the Roche lobe radius, $R/R_{\rm L} = f_\Phi^{-1}$.  The simulated values are compared to the prediction of \citet{1988MNRAS.231..823T}, with three values of the constant $B_{\rm w}$. Unlike the scaling with wind velocity shown in Figure \ref{fig:mdot}, we find that the binary-enhancement in wind mass loss is not single-valued with degree of Roche lobe occupancy, nor is its approximate magnitude described accurately by the \citet{1988MNRAS.231..823T} formula with the default value of $B_{\rm w}=10^4$.  }
\label{fig:mdot}
\end{center}
\end{figure}

One of the previously-applied predictions for tidal enhancement of wind mass loss was postulated by \citet[][their equation 2]{1988MNRAS.231..823T}. This model predicts an enhancement factor depending on the degree of Roche lobe filling, according to
\beq
\frac{\dot M}{\dot M_{\rm hi}}  = 1 + B_{\rm w} \left( \min \left[   \frac{R}{R_{\rm L}} , \frac{1}{2} \right] \right)^6,
\eeq 
where $R/R_{\rm L}$ is the ratio of the stellar radius to the radius of the Roche lobe and $B_{\rm w}$ is a constant with nominal value $10^4$.         
By comparison to our simulation results, we see that the enhancement due to the presence of the binary is less than predicted by \citet{1988MNRAS.231..823T}, and that our simulation results are multi-valued at a given degree of Roche lobe filling. We therefore argue that wind velocity compared to orbital velocity, rather than Roche lobe occupancy, is the most useful determinant of the  rate of mass loss.

\subsection{Angular Momentum Loss Rate}

\subsubsection{Simulated Loss Rates}
We analyze angular momentum carried by the wind in terms of the surface integral of equation \eqref{ldoteval}. In this section, we will refer to specific angular momenta in their dimensionless form, $l/l_{\rm bin}$. Much like mass loss rates, we argue that important trends emerge as a function of wind velocity.

\begin{figure}[tbp]
\begin{center}
\includegraphics[width=0.49\textwidth]{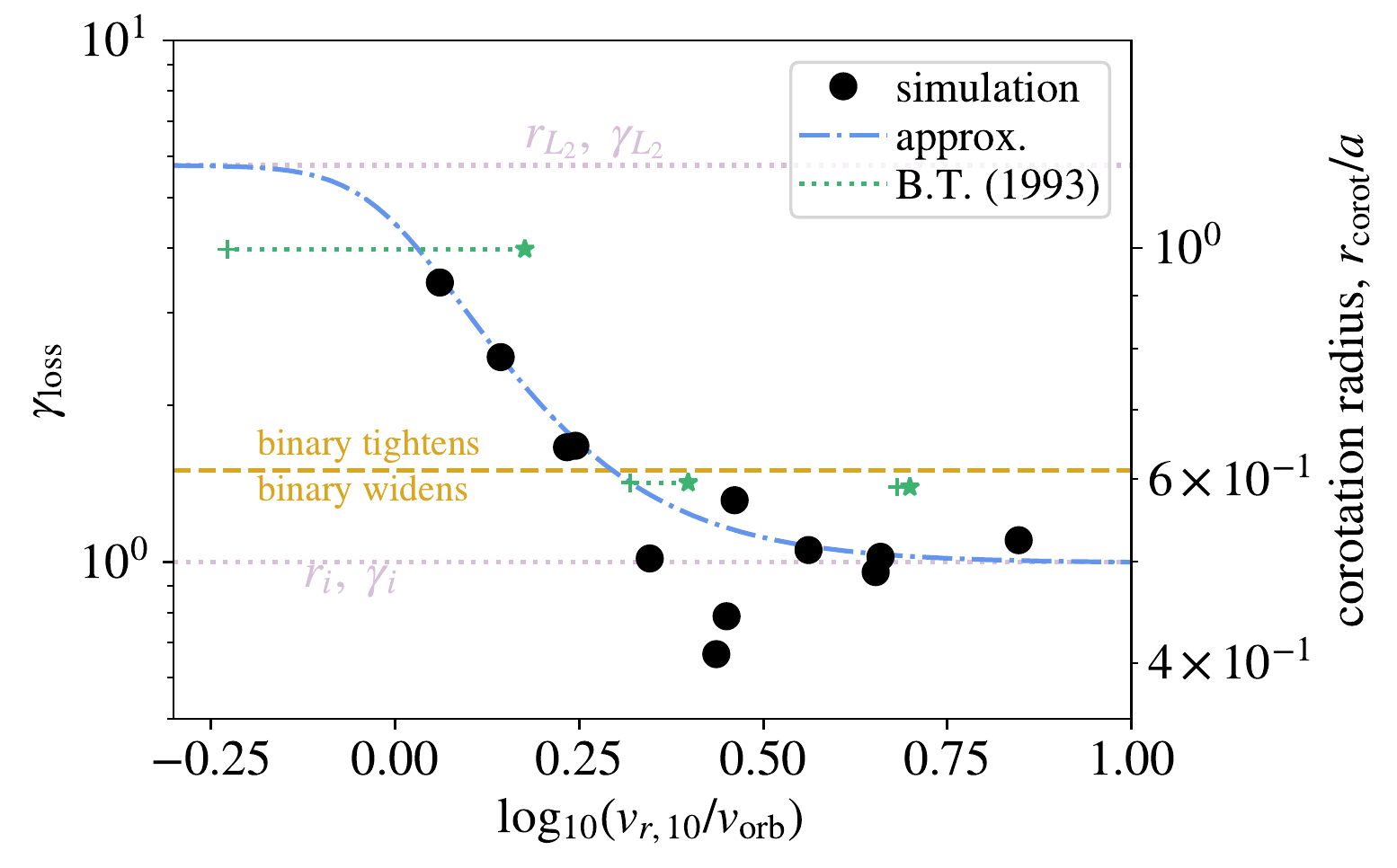}
\caption{The dimensionless specific angular momentum carried by the wind, $\gloss = l_{\rm loss}/l_{\rm bin}$, as a function of wind velocity, $v_{r,10}$. The right-hand axis expresses $\gloss$ in terms of the radius of corotation, $r_{\rm corot}^2 = l_{\rm loss} /\Omega = \gloss l_{\rm bin}/\Omega$, where $\Omega$ is the orbital frequency.  A contour is plotted at the critical value of $\gloss = 3/2$, which separates orbital evolution in which $\dot a >0$ or $\dot a <0$.  High-velocity winds escape with $\gloss \sim 1$. Slower velocity winds lead to $\gloss > 1$ as the winds superimpose and interact with the binary effective potential. We compare to an approximating form, equation \eqref{approxrcorot}, labeled ``approx", and to the results of \citet{1993ApJ...410..719B} for collisionless mass loss, labeled ``B.T. (1993)". }
\label{fig:gloss}
\end{center}
\end{figure}

Figure \ref{fig:gloss} shows the dimensionless specific angular momentum of the wind, $\gloss$, and its dependence on  wind velocity. We show an equivalent right-hand axis of the corotation radius implied by a given angular momentum, where 
\beq
r_{\rm corot} = \sqrt{\frac{l_{\rm loss}}{\Omega} },
\eeq
where $l_{\rm loss}$ is the angular momentum of the wind and $\Omega$ is the orbital frequency.  Figure \ref{fig:gloss} demonstrates that angular momentum losses depend primarily on wind velocity:
\begin{enumerate}[i)]
\item At high wind velocities, winds carry the angular momentum of the binary components, thus $\gloss = \gamma_i = 1$ and $r_{\rm corot} = r_i = a/2$. This high-velocity limit is well-known case of an essentially non-interacting wind that freely escapes, and is sometimes called ``Jeans" mass loss. 
\item In the opposite limit of low-velocity winds, material could be in corotation with the binary all the way out to the radius of the outer Lagrange point, $r_{L_2} \approx 1.2a$. This yields $\gL = r_{L_2}^2 \Omega / l_{\rm bin} \approx 5.76$. Given the shape of the effective potential outside of the outer Lagrange points, there is no mechanism (for a non-magnetic wind) that would maintain corotation to larger radii. We therefore label this upper limit with $r_{L_2}$, $\gL$ in Figure \ref{fig:gloss}. 
\end{enumerate}
In between these limits lies the critical value of $\gloss=3/2$ that separates orbital evolution in which the binary widens due to mass loss ($\gloss < 3/2$) or tightens due to mass loss ($\gloss >3/2$). 

Between the high and low velocity limits of $\gloss$, we must consider the effects of the extended surfaces of the binary components, hydrodynamic stresses from the superposition of winds, and the binary's gravitational torques on the outflowing material. That these overall processes reduce to a nearly one-dimensional function of wind velocity indicates that the primary physical effect must be the expansion velocity of the wind relative to the binary orbital velocity.  Through studying our simulation snapshots, we observe that the wind is maintained in near-corotation out to approximately the sonic surface, along the simulation $x$-axis that connects the binary components. This is visualized most clearly in the snapshots of Figure \ref{fig:machEJ}. Thus, $r_{\rm corot} \sim r_i + Gm_i/(2 c_{\rm s,s}^2) =  r_i + GM/(4 c_{\rm s,s}^2)$ in the intermediate and high-velocity regimes, with a maximal limit at $r_{L_2}$. 

Drawing on this functional form for inspiration, we develop the following approximating formula in terms of the more general property of wind velocity,
\beq\label{approxrcorot}
r_{\rm corot} \approx \left[  r_{L_2}^{-b} + \left(r_i^a +  \left[\frac{GM}{v_{r,10}^2}\right]^a   \right)^{-\frac{b}{a}}\right]^{-\frac{1}{b}},
\eeq
where the associated dimensionless specific angular momentum is 
\beq
\gloss \approx r_{\rm corot}^2 \Omega / l_{\rm bin}. 
\eeq This expression provides a reasonable description of our simulation results across the various wind velocity regimes. We find that $a=1.5$ and $b=5$ fit the simulation data reasonably well. This curve is reproduced with a dot-dash line in Figure \ref{fig:gloss}, where it is labeled ``approx."

\subsubsection{Comparison to Collisionless Mass Loss}

In Figure \ref{fig:gloss} we compare our results for $\gloss$ to those derived by \citet{1993ApJ...410..719B} under a fairly different set of model assumptions. \citet{1993ApJ...410..719B} model a collisionless wind by integrating the ballistic trajectories of particles in the binary potential. This is conceptually similar to but more comprehensive in parameter coverage than earlier work by \citet{1977MNRAS.179..265L}. The wind emerges from one component of the binary system, and \citet{1993ApJ...410..719B} inject their wind with a radial velocity, which in their notation is written $V$ and is normalized to the orbital velocity of the star losing mass. Thus, in our units the injection velocity is $v_{\rm in,BT} = V/2$ for $q=1$. They denote the dimensionless specific angular momentum of particles in the wind as $h_{\rm cm}$, which can be converted to our notation (for $q=1$) by $\gloss = 2h_{\rm cm}$. Finally, for comparison to our results measuring the wind velocity at $r=10$, we compute what the ballistic wind velocity would be, assuming expansion from the single, wind-losing binary component 
\beq
v_{\rm 10,BT} \approx \sqrt{v_{\rm in,BT}^2 - 2 G M_\ast \left( \frac{1}{R_\ast} - \frac{1}{r} \right) }  \approx \sqrt{v_{\rm in,BT}^2 - 1.9} ,
\eeq
where the numerical result comes from $r=10$, $G M_\ast = 0.5$, and $R_\ast = 0.5$.  Because neither of these velocities is completely comparable to those in our hydrodynamic simulations (the velocity profile of decelerating ballistic particles is quite different from that of our accelerating wind), we show both  $v_{\rm in,BT}$ with stars, and $v_{\rm 10,BT}$ with plus symbols in Figure \ref{fig:gloss}. 

Despite the differences in model, we observe a largely similar trend in the results of \citet{1993ApJ...410..719B} and our hydrodynamic simulations. For $v_{r,10} < 2$, we find $\gloss > 1$, while for $v_{r,10} \gtrsim 2$, $\gloss$ asymptotes (in the case of \citet{1993ApJ...410..719B} to $\gloss \approx 1.39$ versus $\gloss \approx 1$ in our simulations). The physical origin of these similar results is somewhat different. In the case of \citet{1993ApJ...410..719B}, the  particles acquire angular momentum (beyond that with which they are injected) through gravitational stresses alone, while in our simulations, hydrodynamic stresses play a role in imparting angular momentum to escaping wind. \citet{1993ApJ...410..719B} do note, however, that in the cases of low-velocity wind only the particles injected on the outer edges of the binary are lost. In our simulations, pressure gradients play a similar regulatory role in allowing material to circulate toward the outer Lagrange points $L_2$ and $L_3$ before it is carried away.

\subsubsection{Origin of Angular Momentum in Interacting Winds}

\begin{figure}[tbp]
\begin{center}
\includegraphics[width=0.49\textwidth]{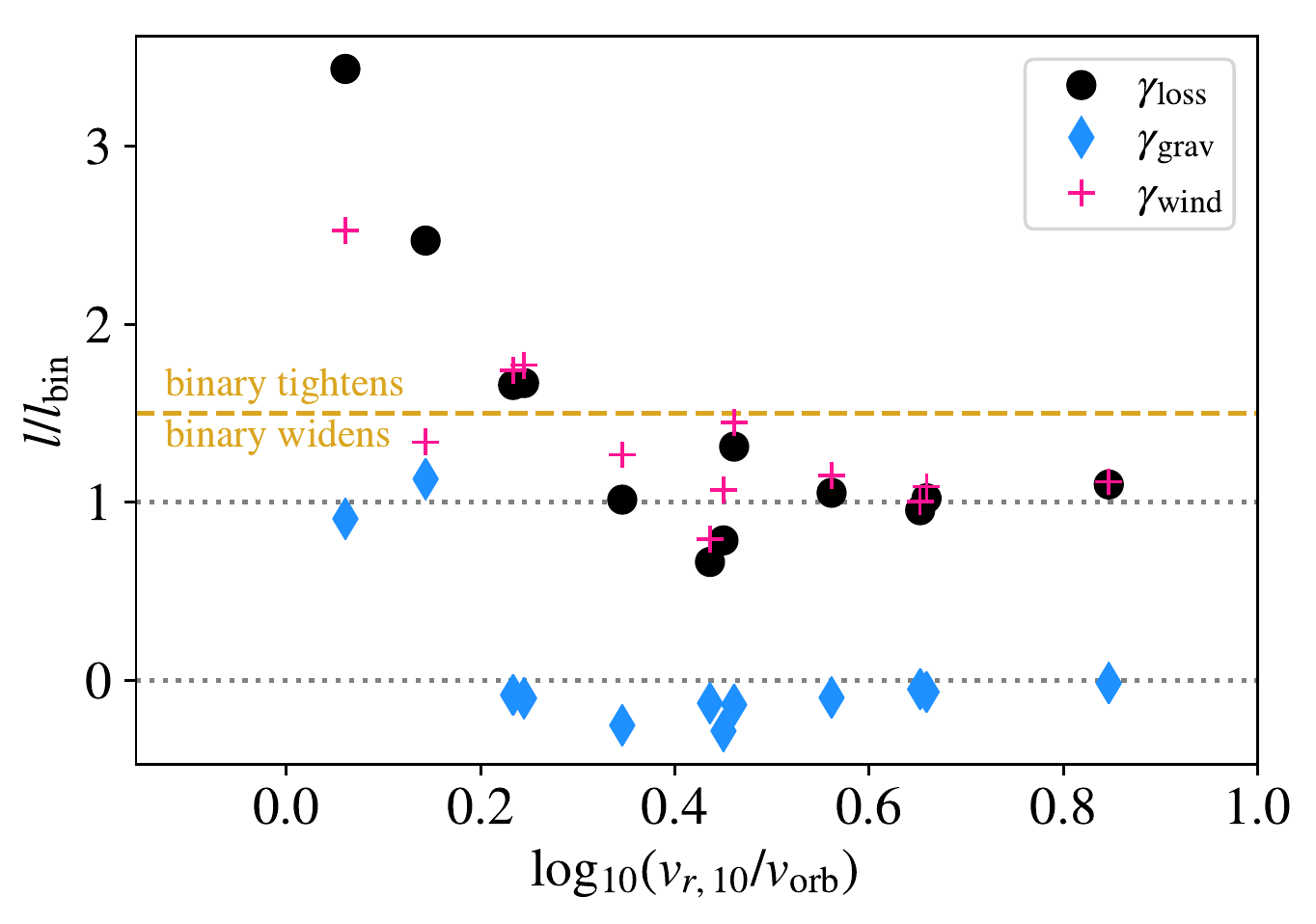}
\caption{Dimensionless angular momenta of the wind resulting from hydrodynamic stresses ($\gw$) or gravitational stresses ($\ggrav$). Together these components comprise the total dimensionless angular momentum, $\gloss$.  All plotted in terms of wind velocity, approximated by $v_{r,10}$. We find that $\gw >0$ in all cases, and is largest for low $f_\Phi$ and high $\lambda$, which lead to low wind velocities. The gravitational addition to $l$ can be either positive or negative, but has smaller magnitude that $\gw$.  }
\label{fig:gwg}
\end{center}
\end{figure}

In Figure \ref{fig:gwg}, we decompose $\gloss$ into the contributing components resulting from gravitational stress ($\ggrav$) or from hydrodynamic stress ($\gw$), equations \eqref{ggrav} and \eqref{gw}, respectively.  This decomposition allows us to trace the origin of angular momentum acquired by the wind as it flows away from the binary. Hydrodynamic stresses include surface forces from the imposition of constant pressure and density along the equipotential surface defined by $f_\Phi$, equation \eqref{PhiSurf}, and the subsequent collisional nature of the gas as it expands away from the binary components and the two, initial separate, winds superimpose. Gravitational stresses arise from net torques on the wind from the binary's gravity  and the resulting exchange between wind and orbital angular momentum. For example, leading edge overdensities in the flow torque the binary forward in its orbit, while trailing overdensities torque the binary backward.

Figure \ref{fig:gwg} reveals that $\gw$ is always greater than zero. It is near unity at high wind velocities, where winds establish separately around each binary component before superimposing. At lower wind velocities, $\gw$ takes on values significantly larger than unity, driving the bulk of the increase in $\gloss$ that we have noted with decreasing velocity in Figure \ref{fig:gloss}. The expanded subsonic region surrounding the pairs at lower wind velocities contributes to an extended region in which pressure forces act to redistribute flow away from the overconcentration near the binary center of mass, imparting it with additional angular momentum. The extended surface area of the binary components at lower $f_\Phi$ also contributes to this addition of angular momentum in the these cases, as the wind can be seen to ``slide" off of the leading edges of the orbiting stars. 

Gravitational stresses as measured by $\ggrav$ in Figure \ref{fig:gwg} approach zero at high wind velocities -- the wind is essentially spherical and non-interacting. At intermediate velocities $\ggrav$ is negative, implying that gravitational stresses remove angular momentum from the wind, adding it to the binary. At low velocities, $\ggrav>0$, and the binary potential imparts angular momentum to the slowly-expanding wind.   At intermediate wind velocities, $2.5< v_{r,10}/v_{\rm orb} < 3$, we observe the widest range of $\gloss$ at similar velocity values. The decomposition by component is useful in disentangling this feature. The highest value of $\gloss$ in this regime comes from model A, for which $f_\Phi=1$. The Roche-lobe filling surfaces of this binary contributes to a high $\gw$, which is slightly reduced by the negative $\ggrav$. The lowest value in this range comes from model L, with $f_\Phi=4$. Here $\gw = 0.79$, but the eventual value of $\gloss$ is significantly lower, $\gloss = 0.67$  due to the negative contribution of $\ggrav$. A similar circumstance occurs with model  G, in which $\gw >1$ but $\gloss<1$ due to a negative $\ggrav$. Thus, while the high-velocity asymptote of the wind specific angular momentum is $\gloss \rightarrow 1$, in the intermediate wind velocity regime, there appears to be some variation at fixed wind velocity depending on other binary parameters, and values of $\gloss < 1$ are possible in this regime.

\section{Discussion}\label{sec:discussion}

\subsection{Limitations \& Astrophysical Applicability}

We have studied a highly idealized numerical problem, in which, among other assumptions, a perfectly symmetric pair of stars, synchronously rotating with their orbit, develop equal winds from high-temperature surfaces. We adopt a nearly-isothermal equation of state for the wind thermodynamics. The resulting hydrodynamic winds accelerate through a sonic point as they expand away from the binary. This model is an extension of one of the simplest models of the solar wind, the \citet{1958ApJ...128..664P} wind model, though the details are slightly different for $\gamma=1.01$, as described in Appendix \ref{sec:wind}. 

Depending on stellar surface properties, one of several different wind acceleration mechanisms is more likely to contribute to the phases when stars lose the most mass. In low-temperature stellar surfaces, radiation pressure on dust that forms within the cooling and expanding wind is a significant wind driving mechanism. Because the radiative flux scales with $r^{-2}$, in its simplest form this sort of wind driving mechanism has the effect of reducing the gravitational attraction of the stars by a factor (which in general is not spatially constant because the formation of dust highly effects the opacity and thus the Eddington ratio as a function of radius). In higher-temperature stellar surfaces, the primary wind driving mechanism is radiation coupling to doppler-broadened metal absorption lines. A similarity is that these massive stars can have luminosities approaching the Eddington luminosity for electron scattering, and thus have reduced effective gravity as well. In the case of a line-driven wind, the resulting wind acceleration profile differs somewhat from that of a hydrodynamic or dust-driven wind \citep{1999isw..book.....L}. 

What each wind model shares is the acceleration of wind from near rest at the stellar surface to some asymptotic velocity at large radii. When the binary system has separation comparable to the characteristic  radius of acceleration, interaction with the binary potential is especially important \citep[e.g.][]{2017MNRAS.468.4465C,2020A&A...637A..91E}. In the case of multiple, radiation driven winds, the superposition of radiative forces and the winds themselves are also important \citep{1996ApJ...469..729C,2009MNRAS.396.1743P,2012A&A...546A..60L,2018MNRAS.477.5640P}. Future work could expand on the simplest-case scenario that we have adopted in this paper to explore how significant these various wind acceleration mechanisms are in determining the morphology of angular momentum carried by the wind from twin-star binaries. 

Although a final conclusion awaits these further studies, we hypothesize that the velocity--mass-loss and velocity--angular-momentum connections established in Figures \ref{fig:mdot} and \ref{fig:gloss} will be relatively robust regardless of the particulars of the wind acceleration process, given that they can be expressed in terms of the relatively general property of wind velocity.  At a minimum, these models are likely more appropriate than assuming that the fast wind approximation of unmodified mass loss rate at $\gloss \sim 1$ holds regardless of binary  properties. Finally, we note that equations $\eqref{binenhance}$ and $\eqref{approxrcorot}$ provide convenient approximations of our simulation results that can be applied on the basis of the ratio of the wind velocity to the orbital velocity, a property that is simple to estimate in binary population models to aid the application of our simulation results to real systems.  

\subsection{Orbital Evolution of Close Twin Binaries}

\begin{figure}[tbp]
\begin{center}
\includegraphics[width=0.45\textwidth]{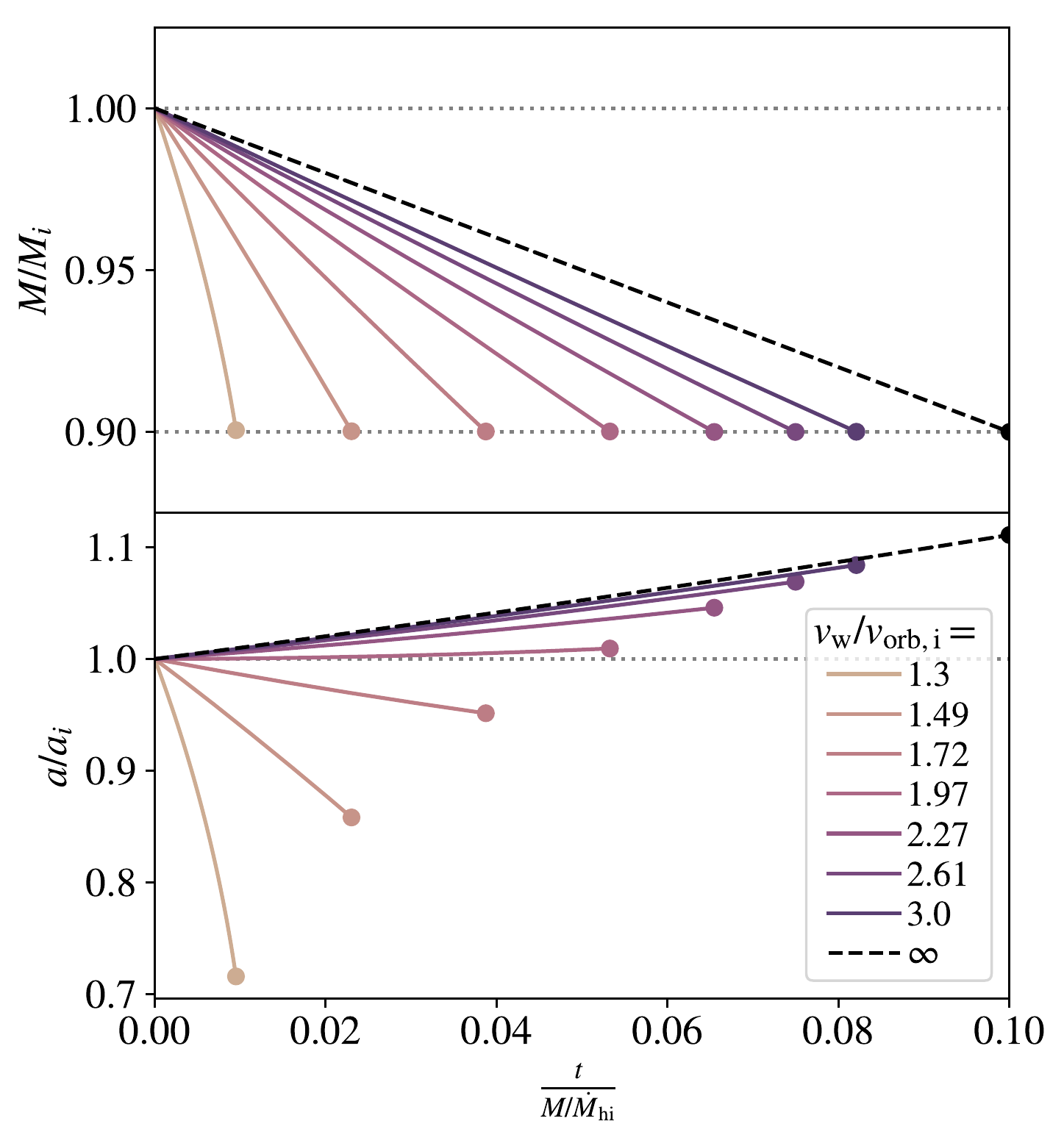}
\caption{Binary mass and separation as a function of time, given different initial ratios of wind velocity to orbital velocity. In each case we consider the range of times during which mass decreases to 90\% of its original value. For lower wind velocities, the mass loss rate is enhanced and the orbit tightens in response.   }
\label{fig:evol}
\end{center}
\end{figure}

To illustrate the implications of our findings for close twin binaries, we compute the orbital evolution of a stellar pair of mass $M$ and separation $a$, with initial mass $M_i$ and separation $a_i$. We assume a constant wind speed, $v_{\rm w}$, and we associate this wind speed with our measured wind speeds at $r=10a$, $v_{r,10}$. We then apply the approximating forms of equations \eqref{binenhance} and \eqref{approxrcorot} to compute the resulting orbital evolution as a function of $\gloss$, following equation \eqref{adot}.  Binaries with high-velocity winds expand, as expected with Jeans mass loss of high-velocity winds. Binaries with lower velocity winds lose mass more rapidly and also contract, in the most extreme cases quite severely. 
 
A key application of these results may be to the chemically-homogeneous evolutionary model for the formation of merging binary black holes \citep{2016MNRAS.458.2634M,2016MNRAS.460.3545D,2016A&A...588A..50M,2016A&A...585A.120S}. In this model, a near-contact massive binary evolves under the influence of tidal stresses to acquire additional rotational mixing such that all or nearly all of each star eventually collapses to a black hole \citep{2009A&A...497..243D}. One such massive contact binary has been observed in 30 Doradus, VFTS 352, which is a near-twin binary with total mass of approximately $59M_\odot$ \citep{2015ApJ...812..102A}.  In many cases, the resulting black hole pairs are close enough to merge in less than the age of the universe \citep[for one example, see Figure 4 of][]{2016A&A...588A..50M}. 

Throughout their evolution, binaries undergoing chemically homogeneous evolution lose a portion of their mass to winds. To give a specific example, the exemplary system evolved in Figure 2 of \citet{2016A&A...588A..50M} loses 10\% of its mass on the main sequence and 20\% prior to the collapse at metallicity of $Z_\odot / 50$.    Because winds are largely metal-line driven, this fraction is thought to be metalicity dependent \citep[e.g.][]{2008A&ARv..16..209P}.  The impact of this wind mass loss on the binary orbits may have important implications for the separation of the binary system. If the separation widens too far, a pair might separate far enough that their remnant black holes would  not merge under the influence of gravitational radiation. On the other hand, if the separation tightens too dramatically, a pair of stars might merge and produce a single, more massive black hole remnant. 

\citet{2016MNRAS.458.2634M} and \citet{2016MNRAS.460.3545D} consider a number of variations of the wind mass loss model, and demonstrate that this is a central parameter in determining the number of merging systems that are observable by LIGO. In one of their model variations, winds are assumed to carry the angular momentum predicted by the model of \citet{1993ApJ...410..719B}. The assumption of $\gloss > 1$ yields their highest predicted observable merger rates at LIGO design sensitivity \citep[see Table 1 of][]{2016MNRAS.460.3545D}. In a simple sense, this is because any changes in binary separation are amplified by the gravitational-wave merger time that scales as $a^4$. By contrast, most population models, such as BSE \citep{2002MNRAS.329..897H}, the chemically homogenous models of \citet{2016A&A...588A..50M},  MOBSE \citep{2018MNRAS.474.2959G}, and SEVN \citep{2019MNRAS.485..889S}, currently adopt the fast-wind approximation of $\gloss =1$ for lack of a complete and more sophisticated prescription. 

Our results suggest that the relative magnitude of the wind speed to the orbital velocity may be a simple representative parameter that determines the binary enhancement in mass loss rate as well as the angular momentum carried away from the binary with the wind (see also \citet{1993ApJ...410..719B}, \cite{ 2005A&A...441..589J}, and \citet{2017MNRAS.468.4465C,2018MNRAS.473..747C}).  Population models incorporating these approximations may be useful step toward understanding how initial properties of close binaries map through their main sequence evolution and toward the formation of binary black holes that may eventually merge.  Indeed, the similarity between our model results and those of \citet{1993ApJ...410..719B} are suggestive of the high predicted black hole merger rates observable by the LIGO-Virgo network (1200~yr$^{-1}$ detections at design sensitivity) under this model variation of \citet{2016MNRAS.458.2634M} and \citet{2016MNRAS.460.3545D}. 

\section{Summary \& Conclusion}\label{sec:conclusion}

We have created and analyzed models of thermal winds from twin-star close binaries in circular orbits. Each of our models adopts $q=1$, and we set the binary components surface based on a factor times the potential at the $L_1$ Lagrange point, $f_\Phi$. The wind initial sound speed is set  based on a hydrodynamic escape parameter, $\lambda$. We examine the emergent wind distributions in each model, as well as the fluxes of mass and angular momentum carried away from the binary by the wind. 

Some key findings of our study are:
\begin{enumerate}
\item The acceleration of winds from about the binary components leads to a symmetric outflow traced by spiral density waves in the plane of the orbit (Figures \ref{fig:rho}). The subsonic acceleration region can surround either the individual binary components, or the entire binary, when $f_\Phi \lesssim \lambda / 2$ (Figure \ref{fig:machEJ}). The resulting outflow is largely similar to that of a spherical wind, with the addition of density and velocity waves in the equatorial plane due to the binary's motion (Figure \ref{fig:rho}).
\item Eventual wind radial velocities are related to the surface sound speeds. At low sound speeds, radial velocities exceed that of a single-object wind model, implying that orbital motion imparts additional kinetic energy to the wind (Figures \ref{fig:vr} and \ref{fig:v10}). 
\item Mass loss rates from the binary are enhanced relative to the estimated superposition of two single-object winds. This effect is best modeled as a function of wind velocity, as shown in Figure \ref{fig:mdot}, and approximated by equation \eqref{binenhance}. 
\item The specific angular momentum carried by the wind depends primarily on the wind velocity as well (Figure \ref{fig:gloss}). It is enhanced well beyond the binary's specific angular momentum for the slower winds, which circulate around the binary before escaping near the outer Lagrange points (Figure \ref{fig:gwg}), and we provide an approximating formula in equation \eqref{approxrcorot}.
\end{enumerate}

The dependence of wind mass loss rate and specific angular momentum on wind velocity, as modeled by equations \eqref{binenhance} and \eqref{approxrcorot} may have implications for models of massive, near contact twin binaries undergoing chemically-homogenous evolution due to rotationally-enhanced mixing. In particular, \citet{2016MNRAS.458.2634M} and \citet{2016MNRAS.460.3545D} argue that the model is sensitive to both the mass and angular momentum carried by winds. Our results suggest that for contact systems with low-velocity winds, mass loss rates and angular momentum loss rates will both be enhanced. Comparison to the model variations of \citet{2016MNRAS.458.2634M} and \citet{2016MNRAS.460.3545D} suggest that the high angular momentum loss rates may indicate an enhancement in the portion of binaries that undergo chemically-homogenous evolution that leave binary black hole remnants that can merge in a Hubble time. Productive avenues for future study include modeling twin-star winds that are radiatively, rather than thermally, accelerated, and applying findings of mass and angular momentum loss rates in population studies of detectable binary black hole mergers with the LIGO-Virgo network.

\acknowledgements{
M.M. is grateful to A. Oklop\v{c}i\'{c} for many discussions of hydrodynamic winds, and to I. Mandel and S. de Mink for conversations about massive, contact binaries that together inspired this work. 
This work was supported by the National Science Foundation under Grant No. 1909203. 
Resources supporting this work were provided by the NASA High-End Computing (HEC) Program through the NASA Advanced Supercomputing (NAS) Division at Ames Research Center. 
This work used the Extreme Science and Engineering Discovery Environment (XSEDE), which is supported by National Science Foundation grant number ACI-1548562. In particular, use of XSEDE resource Stampede2 at TACC through allocation TG-AST190046 enabled this work. 
}

\software{IPython \citep{PER-GRA:2007}; SciPy \citep{2020SciPy-NMeth};  NumPy \citep{van2011numpy};  matplotlib \citep{Hunter:2007}; Astropy \citep{2013A&A...558A..33A}; Athena++ \citep{2020ApJS..249....4S}; XSEDE \citep{xsede} }

\appendix

\section{Spherical Polytropic Winds}\label{sec:wind}
 
 Spherically symmetric polytropic winds serve as a useful benchmark against which to compare the binary wind. We consider a wind that satisfies the mass continuity equation
 \beq
 \dot M = 4 \pi r^2 \rho v
 \eeq
 and has a polytropic equation of state
 \beq
 P = K \rho^\gamma
 \eeq
 The Bernoulli parameter is preserved along a streamline
 \beq
 {\cal B} = \frac{v^2}{2} - \frac{GM}{r} + \frac{\gamma}{\gamma-1}\frac{P}{\rho}
 \eeq
 where the potential is $-GM/r$. 
 From the surface condition at the launching of the wind, we can solve for $K = P_{\rm s}/\rho_{\rm s}^{\gamma}$. Solving the continuity equation for $\rho$, we can rewrite the Bernoulli equation as
\beq
 {\cal B} = \frac{v^2}{2} - \frac{GM}{r} + \frac{\gamma}{\gamma-1} K \left( \frac{\dot M}{4 \pi r^2 v}\right)^{\gamma -1 }
\eeq
 with known initial conditions, the constants $B$, $K$ are known. If we select a value for $\dot M$, we can numerically solve for the full solution for the velocity profile $v(r)$.

\section{Estimates of Mass Loss Rate}\label{sec:mdotest}

\subsection{Single-Object Regime}

We begin with the derivation of the wind mass loss rate from a single object of mass $m$. We assume an isothermal wind, which is similar to, but not identical to, the $\gamma_{\rm ad} =1.01$ model applied in our hydrodynamic simulations. 
In this case, we use the sonic point, at which the wind radial Mach number is unity, to anchor our solution. Thus, $\dot m \approx - \pi r_{\rm sonic}^2 \rho_{\rm sonic} v_{\rm sonic}$. The velocity is equal to the isothermal sound speed, $v_{\rm sonic} = c_{\rm s}$ of the isothermal gas. We can estimate the sonic radius as, 
\beq\label{rsonic}
r_{\rm sonic} = \frac{Gm}{2 c_{\rm s}^2}. 
\eeq
Then, we need to estimate the density at the sonic point. In the subsonic, quasi-hydrostatic region, the density follows exponential decay, so we use that solution to relate the sonic-point density to the surface density as,
\beq
\rho_{\rm sonic} = \rhos \exp \left(  \frac{\Phi_{\rm s}}{c_{\rm s}^2} - \frac{\Phi_{\rm c}}{c_{\rm s}^2}  - \frac{1}{2}\right).
\eeq
The first two terms in the exponential express the potential difference between the surface $(\Phi_{\rm s})$ and the sonic point $(\Phi_{\rm s})$, dividing by the sound speed squared yields the number of scale heights in the quasi-hydrostatic solution. The $\exp(-1/2)$  factor comes from solving the Bernoulli equation for the non-zero velocities in the subsonic region and accounts for the lower density realized with $v_{\rm sonic} = c_{\rm s}$ \citep[page 68]{1999isw..book.....L}. We note that $\Phi_{\rm s}/c_{\rm s}^2 = -\lambda$, and following equation \eqref{rsonic}, $-\Phi_{\rm c}/c_{\rm s}^2=2$. Thus, the expression above can be simplified to
\beq
\rho_{\rm sonic} = \rhos \exp \left( \frac{3}{2} - \lambda \right).
\eeq
Thus the estimated mass loss rate is
\beq
\dot m \approx  -\pi \frac{(Gm)^2 }{c_{\rm s}^3} \rho_{\rm s} \exp \left( \frac{3}{2} - \lambda \right).
\eeq

\subsection{Binary Regime}
Here we estimate a mass loss rate from an equal-mass binary of mass $M$, assuming that a subsonic region encloses the binary components (thus wind escapes from the binary, rather than single components). Under these conditions, we need to take into account the binaries effective potential, including the gravity of the two components and the rotating frame in which the wind is launched. This implies that the wind's subsonic quasi-hydrostatic region cannot expand beyond the outer saddle points of the effective potential, $L_2$ and $L_3$. We will, therefore, derive a mass loss rate considering the flow through these outer Lagrange points as $\dot M \approx A_{\rm L_2} \rho_{L_2} v_{L_2}$. Much like flow through the $L_1$ Lagrange point, we assume that material as velocity equal to its sound speed as it crosses the saddle point, thus $v_{L_2} = c_{\rm s}$ \citep{1975ApJ...198..383L,2017ApJ...835..145J}. The area at the outer Lagrange point is estimated by similar analogy to work considering the $L_1$ Lagrange point. The degree to which gas can spread from the precise saddle point is determined by the scale height, thus,
to order of magnitude, 
\beq
A_{L_2} \sim \frac{\pi c_{\rm s}^2 }{\Omega^2},
\eeq
where $\Omega^2 =  GM / a^3$ is the binary orbital frequency. 

The density at the outer Lagrange point is estimated by analogy to the single object case, as 
\begin{align}
\rho_{L_2} &= \rhos \exp \left(  \frac{\Phi_{\rm s}}{c_{\rm s}^2} - \frac{\Phi_{L_2}}{c_{\rm s}^2}  - \frac{1}{2}\right), \\
&=  \rhos \exp \left(  -\lambda - \frac{\Phi_{L_2}}{c_{\rm s}^2}  - \frac{1}{2}\right),
\end{align}
where $\Phi_{L_2}$ is the effective potential evaluated at the outer Lagrange point. For an equal mass binary, $\Phi_{L_2} \approx 0.86424 \Phi_{L_1} = -1.72848 GM/a$. 

Combining these terms yields,
\begin{align}
\dot M &\approx - 2 \frac{\pi c_{\rm s}^3 }{\Omega^2} \rhos \exp \left(  -\lambda - \frac{\Phi_{L_2}}{c_{\rm s}^2}  - \frac{1}{2}\right), \\
&\approx - 2 \frac{\pi c_{\rm s}^3 a^3}{GM} \rhos \exp \left(  -\lambda - \frac{\Phi_{L_2}}{c_{\rm s}^2}  - \frac{1}{2}\right),
\end{align}
where the additional factor of 2 accounts for symmetric loss through two equal outer Lagrange points. This derivation could be generalized to an unequal binary by providing separate estimates of $\rho_{L_2}$ and $\rho_{L_3}$. 

\subsection{Application to Simulation Models}
To apply these estimates to our simulation models, we begin by acknowledging that our simulation is not strictly isothermal. For the purpose of comparison, however, we associate the surface sound speed, $c_{\rm s,s}$, with the isothermal sound speed of the previous subsections. 

Next, we identify the single-object regime as applicable when winds are sufficiently high velocity as to be relatively unaffected by the binary's orbital motion and potential. In this regime, $r_{\rm sonic} \ll a$ (or similarly $c_{\rm s,s} \gg v_{\rm orb}$), and the wind escapes from an effectively-single object prior to interacting with the binary. The binary regime applies under opposite conditions, when $r_{\rm sonic} \gtrsim a$ (or $c_{\rm s,s} \lesssim v_{\rm orb}$).  In the effectively-single regime, we need to account for mass loss from two objects, each of mass $m=M/2$. With these associations, we define
\beq
\mdhi \approx - \frac{\pi}{2} \frac{(GM)^2 }{c_{\rm s}^3} \rho_{\rm s} \exp \left( \frac{3}{2} - \lambda \right),
\eeq
which approximates the binary mass-loss rate in the high wind-velocity regime, and 
\beq
\mdlo \approx- 2 \frac{\pi c_{\rm s}^3 a^3}{GM} \rhos \exp \left(  -\lambda - \frac{\Phi_{L_2}}{c_{\rm s}^2}  - \frac{1}{2}\right),
\eeq
which approximates the binary mass-loss rate in the low wind-velocity regime.

\bibliographystyle{aasjournal}
\bibliography{wind.bib,software.bib}

\end{document}